\documentclass[11pt,a4paper]{article}
 
\usepackage{amsmath,amsthm,amssymb}
\usepackage{graphics,graphicx}
\usepackage{epsfig}
\usepackage{multicol}
\usepackage{color}
\usepackage{mathrsfs}
\makeatletter
\@addtoreset{equation}{section}
\makeatother

\usepackage{braket}

\usepackage{mathbbol}

\begin{document}
\begin{center}
\LARGE{\bf Generalised Uncertainty Relations for Angular Momentum and Spin in Quantum Geometry}
\end{center}

\begin{center}
{\bf Matthew J. Lake,}${}^{a}$\footnote{matthew5@mail.sysu.edu.cn}\large{\bf Marek Miller}${}^{b}$\footnote{m.miller@cent.uw.edu.pl}\large{\bf and Shi-Dong Liang}${}^{a}$\footnote{stslsd@mail.sysu.edu.cn}
\end{center}
\begin{center}
\emph{${}^a$ School of Physics, Sun Yat-Sen University, Guangzhou 510275, China} \\
\emph{${}^b$ Centre of New Technologies, University of Warsaw, \\ S. Banacha 2c, 02-097 Warszawa, Poland}
\vspace{0.1cm}
\end{center}

\abstract{We derive generalised uncertainty relations (GURs) for orbital angular momentum and spin in the recently proposed smeared-space model of quantum geometry. 
The model implements a minimum length and a minimum linear momentum and recovers both the generalised uncertainty principle (GUP) and extended uncertainty principle (EUP), previously proposed in the 
quantum gravity literature, within a single formalism. 
In this paper, we investigate the consequences of these results for particles with extrinsic and intrinsic angular momentum and obtain generalisations of the canonical ${\rm so(3)}$ and ${\rm su(2)}$ algebras. 
We find that, although ${\rm SO(3)}$ symmetry is preserved on three-dimensional slices of an enlarged phase space, corresponding to a superposition of background geometries, individual subcomponents of the generalised generators obey nontrivial subalgebras. 
These give rise to GURs for orbital angular momentum while leaving the canonical commutation relations intact except for a simple rescaling, $\hbar \rightarrow \hbar + \beta$. 
The value of the new parameter, \mbox{$\beta \simeq \hbar \times 10^{-61}$}, is determined by the ratio of the dark energy density to the Planck density, and~its existence is required by the presence of both minimum length and momentum uncertainties.  
Here, we assume the former to be of the order of the Planck length and the latter to be of the order of the de Sitter momentum $\sim \hbar\sqrt{\Lambda}$, where $\Lambda$ is the cosmological constant, which is consistent with the existence of a finite cosmological horizon. 
In the smeared-space model, $\hbar$ and $\beta$ are interpreted as the quantisation scales for matter and geometry, respectively, and~a quantum state vector is associated with the spatial background. 
We show that this also gives rise to a rescaled Lie algebra for generalised spin operators, together with associated subalgebras that are analogous to those for orbital angular momentum. 
Remarkably, consistency of the algebraic structure requires the quantum state associated with a flat background to be fermionic, with spin eigenvalues $\pm \beta/2$.
Finally, the modified spin algebra leads to GURs for spin measurements. 
The potential implications of these results for cosmology and high-energy physics, and~for the description of spin and angular momentum in relativistic theories of quantum gravity, including dark energy, are briefly discussed.}


\tableofcontents

\section{Introduction} \label{Sec.1}

GURs for position and linear momentum are motivated by gedanken experiments in phenomenological quantum gravity \cite{Adler:1999bu,Scardigli:1999jh,Bolen:2004sq,Park:2007az,Bambi:2007ty}. 
An advantage of this approach is that, being based on very general considerations, the resulting phenomenology is expected to be model-independent. 
Generalisations of the Heisenberg uncertainty principle (HUP) are, therefore, a fairly generic prediction of low-energy quantum gravity, no matter how much 
individual models differ in their conceptual bases or mathematical structures. 
GURs for position and momentum have also been motivated by arguments in string theory, non-commutative geometry, and~deformed special relativity, among others~\cite{Tawfik:2015rva,Tawfik:2014zca}. 

A closely related and very general prediction of quantum gravity models is the existence of a minimum length scale, which is expected to be of the order of the Planck length $\sim\sqrt{\hbar G/c^3}$ \cite{Hossenfelder:2012jw,Garay:1994en}. 
Certain models also incorporate a minimum momentum scale, although there is less general agreement about whether this is an essential feature of any would-be quantum gravity theory and, if so, on what value the minimum momentum should take \cite{Kempf:1996ss,Kim:2007hf,Asghari:2013sda,Stetsko:2012kx}. 
Nonetheless, a minimum momentum scale is consistent with known physics as the existence of a positive cosmological constant $\Lambda > 0$, inferred from observations of type 1A supernovae, large-scale structure, and~the cosmic microwave background (CMB) \cite{Aghanim:2018eyx,Betoule:2014frx}, implies a minimum space-time curvature. 
This, in turn, implies a maximum horizon distance of the order of the de Sitter length $\sim 1/\sqrt{\Lambda}$ \cite{Spradlin:2001pw}. 
This is approximately equal to the present day horizon radius, $r_{\rm U} \simeq 10^{28}$ cm \cite{Hobson:2006se}. 
Thus, in a universe with positive scalar curvature of order $\Lambda$, any uncertainty principle with a leading order Heisenberg term gives rise to a minimum momentum of order $\sim \hbar\sqrt{\Lambda}$ since $\Delta x$ is bounded from above by the de Sitter scale.  

Motivated by these considerations, two of the most intensively studied GURs in the quantum gravity literature are the GUP \cite{Adler:1999bu,Scardigli:1999jh}, 
\begin{eqnarray} \label{GUP_rough}
\Delta x \gtrsim \frac{\hbar}{2\Delta p} + \alpha \frac{G \Delta p}{c^3} \, , 
\end{eqnarray}
and the EUP \cite{Bolen:2004sq,Park:2007az,Bambi:2007ty},  
\begin{eqnarray} \label{EUP_rough}
\Delta p \gtrsim \frac{\hbar}{2\Delta x} + \eta \hbar\Lambda \Delta x \, ,
\end{eqnarray}
where $\alpha$ and $\eta$ are numerical constants of order unity.
The former implements a minimum length scale, $\sim\sqrt{\hbar G/c^3}$, but no minimum momentum, and~may be obtained by extending the Heisenberg microscope thought experiment to include the gravitational attraction between the massive particle and the probing photon \cite{Adler:1999bu,Scardigli:1999jh}. 
The latter implements a minimum momentum scale, $\sim \hbar\sqrt{\Lambda}$, but no minimum length, and~may be obtained by modifying Heisenberg's argument to include the effects of repulsive dark energy \cite{Bolen:2004sq,Park:2007az,Bambi:2007ty}. 
Thus, taking the GUP or EUP separately breaks position-momentum symmetry in the uncertainty relations. 
However, taking into account the effects of both canonical gravitational attraction and repulsive dark energy motivates the extended generalised uncertainty principle (EGUP) \cite{Kempf:1996ss}:
\begin{eqnarray} \label{EGUP_rough}
\Delta x \Delta p \gtrsim \frac{\hbar}{2} + \tilde{\alpha} (\Delta p)^2 +  \tilde{\eta} (\Delta x)^2 \, .
\end{eqnarray}    
Here, $\tilde{\alpha}$ and $\tilde{\eta}$ are appropriate dimensionful constants, which may be obtained by comparing Equation~(\ref{EGUP_rough}) with Equations (\ref{GUP_rough}) and (\ref{EUP_rough}), respectively. 
In this way, the GUP and EUP may be obtained as appropriate limits of the EGUP. 

Unfortunately, the greatest strength of the phenomenological approach, namely, its model-independence, is also its greatest drawback. 
Although the HUP may be motivated, heuristically, by Heisenberg's thought experiment, it may also be obtained rigorously from the canonical quantum formalism \cite{Ish95}. 
In the former, $\Delta x$ and $\Delta p$ represent, somewhat vaguely, unavoidable imprecisions in the position and momentum of a quantum particle.  
In the latter, $\Delta_{\psi} x$ and $\Delta_{\psi} p$ represent standard deviations of the probability density associated with the quantum wave function, $|\psi|^2$.
Thus, these quantities denote well defined measures of the width of the wave function in the position and momentum space representations, respectively. 
In this paper, we distinguish between heuristic and well defined uncertainties by labeling the latter with a subscript. 
This indicates the specific wave function from which the underlying probability distribution is derived. 

By contrast with the HUP, there is currently no consensus on how to implement the GUP, EUP or EGUP within a well defined quantum formalism. 
One option, which until recently was the only possibility considered in the existing literature, is to modify the canonical commutation relations for position and momentum \cite{Tawfik:2015rva,Tawfik:2014zca,Hossenfelder:2012jw}. 
Non-canonical terms in the Heisenberg algebra then generate additional terms in the uncertainty relations via the Schr{\" o}dinger--Robertson bound \cite{Ish95}. 
In this scenario, a rigorously defined version of the EGUP (\ref{EGUP_rough}), with the heuristic uncertainties $\Delta x$ and $\Delta p$ replaced by $\Delta_{\psi} x$ and $\Delta_{\psi} p$, respectively, may be obtained by modifying the canonical position and momentum operators such that \cite{Kempf:1996ss}:  
\begin{eqnarray} \label{mod_comm}
[\hat{x},\hat{p}] = i\hbar +  2\tilde{\alpha} \hat{p}^2 +  2\tilde{\eta} \hat{x}^2 \, . 
\end{eqnarray}

Unfortunately, although such approaches yield the expected phenomenology (i.e., minimum-length GURs), they remain plagued by theoretical and mathematical difficulties, even after nearly 25 years of research \cite{Kempf:1994su,Kempf:1996ss}. 
Most notably, it may be shown that modifications of the Heisenberg algebra automatically imply violation of the equivalence principle \cite{Tawfik:2015rva,Tawfik:2014zca}. 
Thus, the ``correct'' quantum gravity phenomenology is obtained at very high price, that is, by sacrificing the founding principle of classical general relativity. 
Even more seriously, modified commutators give rise to the so-called soccer ball problem for multiparticle states, and~it is unclear whether a sensible macroscopic limit can be consistently defined within such a formalism \cite{Hossenfelder:2012jw,Hossenfelder:2014ifa,LakeUkraine2019}. 
 
An alternative approach, recently considered in \cite{Lake:2018zeg}, is to modify the canonical quantum wave function and, hence, the underlying probability distribution from which all operator uncertainties are derived. 
The basic idea, proposed in the so-called smeared-space model, is to associate quantum state vectors with spatial points in the classical background geometry. 
By the principle of quantum superposition, this allows a quantum state to be associated with the background space as a whole. 
The resulting extended phase space may be interpreted as a quantum superposition of geometries, and~the canonical quantum state, $\ket{\psi} \in \mathcal{H}$, is mapped such that $\ket{\psi} \mapsto \ket{\Psi} \in \mathcal{H} \otimes \mathcal{H}$. 
The generalised state vector $\ket{\Psi}$ describes the evolution of quantum matter on a quantum background, incorporating geometric superpositions, and~the associated wave function $\braket{\vec{r},\vec{r}{\, '} | \Psi}$ is given by $\Psi(\vec{r},\vec{r}{\, '}) = \psi(\vec{r})g(\vec{r}{\, '}-\vec{r})$. 
The function $g(\vec{r}{\, '}-\vec{r})$ is known as the ``smearing function'' and has a finite width in both position and momentum space. 
Hence, points in the classical phase space of the system $(\vec{r},\vec{p})$ become "smeared" over finite minimum volumes in the position and momentum space representations of the corresponding quantum theory. 
Unlike in canonical QM, absolute limits are set to each \cite{Lake:2018zeg}. 

Important consequences of the model include the existence of minimum position and momentum uncertainties, leading to rigorously defined analogues of the heuristic GUP, EUP and EGUP relations~(\ref{GUP_rough})--(\ref{EGUP_rough}), and~the emergence of a minimum energy density in nature. 
The latter is an unavoidable consequence of the theory and is of the order of the observed dark energy density {\it if} the minimum smearing scales are chosen to be of the order of the Planck length $\sim \sqrt{\hbar G/c^3}$ and the de Sitter momentum $\sim \hbar\sqrt{\Lambda}$. 
The canonical position and momentum operators must also be modified such that $\hat{x}^i \mapsto \hat{X}^i$ and $\hat{p}_j \mapsto \hat{P}_j$, where $\hat{X}^i$ and $\hat{P}_j$ act on the tensor product Hilbert space. 
However, crucially, the resulting commutation relations are simply a rescaled representation of the Heisenberg algebra, with $\hbar \rightarrow \hbar + \beta$, where $\beta \sim \sqrt{\hbar^3 G \Lambda/c^3}$. 
The model is therefore consistent with the equivalence principle and provides a neat solution of the soccer ball problem that plagues approaches based on modified commutators \cite{LakeUkraine2019,Lake:2018zeg}. 

Here, we extend the analysis presented in \cite{Lake:2018zeg} to include angular momentum and spin. 
(For previous work on the implications of GURs for angular momentum, based on the modified commutator formalism, see \cite{Bosso:2016frs}.)
Although our analysis remains non-relativistic, we note the close connection between the rotation generators in flat space and the Lorentz generators in flat space-time \cite{Peskin:1995ev}. 
Thus, by generalising the operators for non-relativistic angular momentum and spin to incorporate the effects of smeared-space, we aim to lay the foundations for the construction of a relativistic theory of quantum matter evolving on a quantum background geometry. 

The structure of this paper is as follows. 
We begin, in Section \ref{Sec.2}, by discussing important problems faced by any model of angular momentum (and hence spin) in a universe with GURs for position and linear momentum. 
Though crucial, the problems considered appear to have been largely overlooked in the existing literature, so we consider them in detail. 
Section \ref{Sec.3} gives a brief review the of smeared-space formalism originally presented in \cite{Lake:2018zeg}. 
This is intended as a self-contained introduction for readers not familiar with our previous work, so readers with a basic understanding of the model may skip this section.
Our new results are presented in Sections \ref{Sec.4} and \ref{Sec.5}.

The generalised angular momentum operators are derived in Section \ref{Sec.4} where it is shown that each operator may be split into the sum of three terms. 
The first term is a ``pure'' matter piece which acts nontrivially on the first subspace of the tensor product Hilbert space. 
This corresponds to the canonical angular momentum operator and generates one copy of the ${\rm so}(3)$ Lie algebra, scaled by $\hbar$. 
The second is a ``pure'' geometric piece which acts nontrivially on the second subspace. 
This corresponds to the angular momentum of the background and generates an additional copy of the ${\rm so}(3)$ algebra, scaled by $\beta$. 
The third term is the interaction term which determines how the angular momentum of a quantum particle and the angular momentum of the background influence one another.  
The generalised spin operators are derived in Section \ref{Sec.5} and obey a similar three-way split.
Our most important results are the generalised algebras obeyed by individual subcomponents. 
In Sections \ref{Sec.4} and \ref{Sec.5}, it is shown that the unique structure of these algebras gives rise to GURs, for both angular momentum and spin, while leaving the canonical commutators unchanged except for a simple rescaling, $\hbar \rightarrow \hbar + \beta$. 
This is a crucial difference between the smeared-space model and previous approaches presented in the literature. 
Section \ref{Sec.6} contains a summary of our conclusions and a discussion of prospects for future~work.

Throughout this work, we encounter various technical, conceptual and philosophical problems of a somewhat subtle nature. 
These are discussed in greater detail in Appendices \ref{Appendix.A}--\ref{Appendix.D}. 
The interested reader is referred to the relevant appendix at the appropriate point within the main text. 
However, sufficient detail is contained within the main body of the article for it to be read as a self-contained whole, so that the appendices may also be skipped, if desired. 

\section{Conceptual Problems for Angular Momentum and Spin in GUR Scenarios} \label{Sec.2}

As angular momentum is a pseudo-vector, relations involving angular momentum are, inherently, vector relations \cite{LandauMechanics}. 
In canonical QM, this poses no fundamental difficulties: although tangent vectors in the background geometry 
define the classical metric \cite{Nakahara:2003nw,Frankel:1997ec}, this is left unchanged by the quantisation of matter living ``in'' the classical space \cite{Rae}. 
However, in the context of a would-be theory of {\it quantum mechanical space} and hence any would-be theory of {\it quantum gravity}, such relations must be handled with extreme care. 
For this reason, before presenting our main results, we discuss a number of subtle technical and conceptual problems that arise in this context. 
These naturally come to the fore when attempting to construct generalisations of {\it vector} relations to include the effects of quantum fluctuations of the background.

To begin, we note that the canonical angular momentum algebra is obtained by replacing the classical position and momentum values, occurring in the usual expression for the components of the angular momentum vector $l_i = \epsilon_{ij}{}^{k}x^jp_k$ (*), with position and momentum operators satisfying the Heisenberg algebra, $[\hat{x}^j,\hat{p}_k] = i\hbar\delta^{j}{}_{k}$. 
This serves as the definition of the angular momentum operators, $\hat{l}_i :=\epsilon_{ij}{}^{k}\hat{x}^j\hat{p}_k$, and~is sufficient to ensure that the canonical algebra $[\hat{l}_i,\hat{l}_j] = i\hbar\epsilon_{ij}{}^{k}\hat{l}_k$ holds. 
In other words, the definition (*), plus the requirement that the canonical position-momentum commutator also holds, is sufficient to ensure that the angular momentum operators obey the ${\rm so}(3)$ Lie algebra, scaled by $\hbar$ \cite{Rae,Ish95}.

Thus, for models in which GURs for position and momentum are generated by modified commutation relations, it is natural to ask what happens to the corresponding angular momentum operators and their associated algebra. 
Furthermore, given the close connection between orbital angular momentum and spin or, equivalently, between the ${\rm so}(3)$ and ${\rm su}(2)$ Lie algebras \cite{Jones98}, it is natural to ask what implications (if any) such relations have for the intrinsic spin of particles in a quantum gravity scenario. 
Despite this, only a handful of articles in the huge literature on GURs explicitly consider angular momentum and/or spin. 
In addition, many of these simply replace the classical values $x^j$ and $p_k$, occurring in (*), with generalised position and momentum operators satisfying a modified Heisenberg algebra, i.e., $[\hat{X}^j,\hat{P}_k] \neq i\hbar\delta^{j}{}_{k}$. This then serves as the definition of the generalised angular momentum operators, $\hat{L}_i :=\epsilon_{ij}{}^{k}\hat{X}^j\hat{P}_k$, and~induces a modified angular momentum algebra. 
There is, however, a potentially significant problem with this approach that has so far been overlooked in the existing literature. 
We now outline this and consider ways in which it can be overcome in the smeared-space model.

To understand the problem, we first note that the standard expression for the components $l_i$ is derived from the vector relation $\vec{l} = \vec{r} \times \vec{p}$, where $\vec{r}$ is the displacement vector in physical space and $\vec{p}$ is the displacement in momentum space. 
Specifically, in the expression (*) the components $l_i$ are obtained by projecting the vector $\vec{l}$ onto a {\it global Cartesian basis}. 
However, both displacement vectors and the {\it global} Cartesian basis in which they are expressed, e.g., $\vec{r} = x^{i}\underline{{\bf e}}_i(0)$, only exist in flat Euclidean space \cite{Nakahara:2003nw,Frankel:1997ec}.
(Here, ${\bf \underline{e}}_i(0)$ denotes the tangent vector defined at the coordinate origin, which, by convention, is located at the centre of rotation.)  
Furthermore, this space is invariant under global shift isometries, and~the canonical Heisenberg algebra is simply a representation of the shift-isometry algebra, again scaled by $\hbar$ \cite{Ish95}. 
Thus, modifying the Heisenberg algebra is equivalent to modifying the symmetry group of the background space in which the quantum particles ``live''. 
If the generalised position and momentum operators do not satisfy the canonical commutation relation, up to a rescaling by $\hbar$ or another constant factor with dimensions of action, this space cannot be Euclidean, and~the expression $\hat{L}_i = \epsilon_{ij}{}^{k}\hat{X}^j\hat{P}_k$ is invalid.

At first glance, this argument seems rather technical and abstract. 
However, it is not. 
It is in accordance with our intuition about the physical nature of the quantum background space and with the prevailing interpretation of GURs presented in the literature. 
Typically, the minimum resolution implied by GURs is attributed to Planck scale fluctuations of the background geometry, below which the concept of distance becomes ill defined \cite{Hossenfelder:2012jw,Garay:1994en}. 
Fluctuations of the background metric are also expected to generate curvature fluctuations over very small length scales, comparable to the Planck length \cite{Padmanabhan:1985ap,Padmanabhan:1985jq}. 
Thus, if the latter occur, it is clear that space cannot be Euclidean at the Planck scale or on the very small scales over which GUP-type corrections are expected to be relevant \cite{Tawfik:2015rva,Tawfik:2014zca}.

Potentially, this problem also affects the smeared-space model. 
In \cite{Lake:2018zeg} a method for determining the probability amplitude associated with a given metric fluctuation, induced by the coherent transitions allowed by smearing, was outlined. 
This relied on a specific interpretation of the smearing function $g(\vec{r}{\, '}-\vec{r})$ and is incompatible with the expression $\hat{L}_i :=\epsilon_{ij}{}^{k}\hat{X}^j\hat{P}_k$, where $\hat{X}^j$ and $\hat{P}_k$ are simply the smeared position and momentum operators defined previously in \mbox{\cite{Lake:2018zeg}. }
In~other words, when dealing with angular momentum, our model faces the same potential problems as models based on modified commutators---as long as GUP-type corrections are attributed to fluctuations of the background, one must be {\it very} careful when attempting to generalise the {\it vector} relations of canonical QM!

In short, since classical particles ``live'' on a fixed background geometry, one may quantize their dynamics but not the background, by promoting only phase space coordinates to Hermitian operators. 
This leaves the geometric basis vectors to which the coordinate values refer unchanged, e.g., $\vec{r} = x^{i}{\bf \underline{e}}_i(0) \mapsto \hat{\vec{r}} := \hat{x}^{i}{\bf \underline{e}}_i(0)$, and~is a subtle mathematical expression of the ontological split between quantum particles and classical geometry. 
Hence, in any would-be model of a genuine quantum background, it is not {\it automatically} clear that we can consistently substitute generalised operators into the equations of canonical QM. 
These issues are discussed in detail in the Appendices, and~we refer the interested reader to Appendix \ref{Appendix.A} for a more in depth analysis of these subtle but important problems.

Despite this, it is possible {\it with sufficient care} to justify the use of the expression $\hat{L}_i :=\epsilon_{ij}{}^{k}\hat{X}^j\hat{P}_k$ in the smeared-space model. 
To do this, we must ensure that the coherent transitions between points in the background, allowed by smearing, do not change the underlying (flat) nature of the geometry. 
This requires us to take a stricter interpretation of the smearing function, $g(\vec{r}{\, '}-\vec{r})$, than the one originally presented in \cite{Lake:2018zeg}.

To this end, we interpret the extended Hilbert space of the model as representing particle states in a superposition of flat Euclidean geometries. Each possible geometry differs from the ``original'' classical geometry only by a pair-wise exchange of points, $\vec{r} \leftrightarrow \vec{r}{\, '}$, and~$g(\vec{r}{\, '}-\vec{r})$ is interpreted as the probability amplitude for this transition. 
Such a transition is also assumed to exchange the values of the canonical QM wave function associated with each point, i.e., $\psi(\vec{r}) \leftrightarrow \psi(\vec{r}{\, '})$ but cannot change the underlying flatness of the background. 
(The back-reaction of $\psi(\vec{r})$ on the geometry is neglected in this approximation.) 
Overall, this results in additional stochastic fluctuations in the position of a quantum particle, vis-{\`a}-vis the canonical theory, in which the background is fixed and classical. 
This results in a minimum irremovable uncertainty in position, of the order of the width of $g(\vec{r}{\, '}-\vec{r})$.

The basic formalism of the original smeared-space model \cite{Lake:2018zeg}, in which arbitrary transitions of the form $\vec{r} \rightarrow \vec{r}{\, '}$ were permitted (giving rise to metric fluctuations), is summarised in Section \ref{Sec.3}. 
The stricter interpretation of the smearing function, used throughout the rest of this work, is discussed in greater detail in Appendix \ref{Appendix.B}. 
From Section \ref{Sec.3} onwards, the stricter interpretation is employed, so that analogues of the canonical QM equations may be obtained simply by substituting $\hat{x}^i \rightarrow \hat{X}^i$, $\hat{p}_j \rightarrow \hat{P}_j$, etc., where capitals denote the relevant smeared-space operators. 
We stress that the results presented in Sections  \ref{Sec.4} and \ref{Sec.5} are new, whereas those presented in Section \ref{Sec.3} review previously published material.

\section{Recap of the Smeared-Space Model} \label{Sec.3}

In \cite{Lake:2018zeg}, a new model of quantum geometry was proposed in which each point $\vec{r}$ in the classical background is associated with a vector $\ket{g_{\vec{r}}}$ in a Hilbert space, where
\begin{eqnarray} \label{|g_r>}
\ket{g_{\vec{r}}} := \int g(\vec{r}{\, '}-\vec{r}) \ket{\vec{r}{\, '}} {\rm d}^3\vec{r}{\, '} \, ,
\end{eqnarray}
and $g(\vec{r}{\, '}-\vec{r})$ is any normalised function. 
For simplicity, however, we may imagine $g(\vec{r})$ as a Gaussian centred at $\vec{r}=0$. 

Since each point in the background geometry may be associated (heuristically) with a Dirac delta $\delta^3(\vec{r}{\, '}-\vec{r})$ or, equivalently, a ket $\ket{\vec{r}}$, the background space may be "smeared" by mapping each point to a superposition of all points via:
\begin{eqnarray} \label{smear_map}
\ket{\vec{r}} \mapsto \ket{\vec{r}} \otimes \ket{g_{\vec{r}}} \, . 
\end{eqnarray}

We may visualise the smearing map (\ref{smear_map}) as follows: for each point $\vec{r} \in \mathbb{R}^3$ in the classical geometry we obtain one whole ``copy'' of $\mathbb{R}^3$, thus doubling the size of the classical phase space. 
The resulting smeared geometry is represented by a six-dimensional volume, namely, $\mathbb{R}^3 \times \mathbb{R}^3$, in which each point $(\vec{r},\vec{r}{\, '})$ is associated with a complex number $g(\vec{r}{\, '}-\vec{r})$. 
This is interpreted as the quantum amplitude for the transition $\vec{r} \rightarrow \vec{r}{\, '}$, and~the higher-dimensional space is interpreted as a superposition of three-dimensional geometries \cite{Lake:2018zeg}.

Hence, in this model, ``points'' in the background exist in a superposition of states, and~may undergo stochastic fluctuations as the result of measurements. 
This also affects the statistics of the canonical quantum matter living on, or ``in'', the space. 
Specifically, using (\ref{smear_map}), the canonical quantum state $\ket{\psi} = \int \psi(\vec{r}) \ket{\vec{r}}{\rm d}^3\vec{r}$ is mapped according to $\ket{\psi} \mapsto \ket{\Psi}$, where:    
\begin{eqnarray} \label{Psi_x}
\ket{\Psi} := \int\int \psi(\vec{r})g(\vec{r}{\, '}-\vec{r}) \ket{\vec{r},\vec{r}{\, '}} {\rm d}^{3}\vec{r}{\rm d}^{3}\vec{r}{\, '} \, . 
\end{eqnarray}
The corresponding expansion in smeared momentum space is given by: 
\begin{eqnarray} \label{Psi_p}
\ket{\Psi} := \int\int \psi_{\hbar}(\vec{p})\tilde{g}_{\beta}(\vec{p}{\, '}-\vec{p}) \ket{\vec{p} \, \vec{p}{\, '}} {\rm d}^{3}\vec{p}{\rm d}^{3}\vec{p}{\, '} \, ,
\end{eqnarray}
where 
\begin{eqnarray} \label{dB-1}
\psi_{\hbar}(\vec{p}) = \frac{1}{\sqrt{2\pi\hbar}} \int \psi(\vec{r}) e^{-\frac{i}{\hbar}\vec{p}.\vec{r}}{\rm d}^{3}\vec{r} \, , 
\end{eqnarray}
and
\begin{eqnarray} \label{}
\tilde{g}_{\beta}(\vec{p}{\, '}-\vec{p}) := \frac{1}{\sqrt{2\pi\beta}} \int g(\vec{r}{\, '}-\vec{r}) e^{-\frac{i}{\beta}(\vec{p}{\, '}-\vec{p}).(\vec{r}{\, '}-\vec{r})} {\rm d}^{3}\vec{r}{\rm d}^{3}\vec{r}{\, '} 
\end{eqnarray}
with $\beta \neq \hbar$ \cite{Lake:2018zeg}. 

In other words, the momentum space representation of the canonical quantum wave function $\psi_{\hbar}(\vec{p})$ is given by the Fourier transform of $\psi(\vec{x})$, which is transformed at the scale $\hbar$. 
(Here, we use the subscript $\hbar$ to emphasise this point.) 
By contrast, the momentum space representation of the geometric part of the composite quantum state $\ket{\Psi}$, $\tilde{g}_{\beta}(\vec{p}{\, '}-\vec{p})$, is given by the Fourier transform of $g(\vec{r}{\, '}-\vec{r})$, where the transformation is performed at a new scale $\beta$. 
This represents the quantisation scale for space (rather than matter) and must be fixed by physical considerations. 
In \cite{Lake:2018zeg}, it was shown that, in order to reproduce the observed vacuum energy density, $\rho_{\Lambda} = \Lambda c^2/(8\pi G) \simeq 10^{-30} \, {\rm g \, . \, cm^{-3}}$, where $\Lambda \simeq 10^{-56} \, {\rm cm^{-2}}$ is the cosmological constant \cite{Hobson:2006se}, $\beta$ must take the order of magnitude value:
\begin{eqnarray} \label{beta_mag}
\beta = 2\hbar\sqrt{\frac{\rho_{\Lambda}}{\rho_{\rm Pl}}} \simeq \hbar \times 10^{-61} \, , 
\end{eqnarray}
where $\rho_{\rm Pl} \simeq 10^{93}  \, {\rm g \, . \, cm^{-3}}$ is the Planck density. 

Consistency between Equations (\ref{Psi_x}) and (\ref{Psi_p}) requires:
\begin{eqnarray} \label{mod_dB}
\braket{\vec{r} , \vec{r}{\, '}|\vec{p} \, \vec{p}{\, '}} := \frac{1}{2\pi\sqrt{\hbar\beta}} e^{\frac{i}{\hbar}\vec{p}.\vec{r}} e^{\frac{i}{\beta}(\vec{p}{\, '}-\vec{p}).(\vec{r}{\, '}-\vec{r})} \, .
\end{eqnarray}
Hence, $\ket{\vec{p} \, \vec{p}{\, '}}$ represents an entangled state in the rigged basis of the ``enlarged'' Hilbert space $\mathcal{H} \otimes \mathcal{H}$, where $\mathcal{H}$ is the Hilbert space of canonical (three-dimensional) QM. \footnote{Strictly, the Hilbert space is not enlarged, since $\mathcal{H} \otimes \mathcal{H} \cong \mathcal{H}$ when $\mathcal{H}$ denotes the Hilbert space with countably infinite dimensions, i.e., the Hilbert space of canonical QM in {\it any} number of (physical) spatial dimensions \cite{HilbertSpaces}.} 
We emphasise this by not writing a comma between $\vec{p}$ and $\vec{p}{\, '}$, by contrast with $\ket{\vec{r},\vec{r}{\, '}} := \ket{\vec{r}}\ket{\vec{r}{\, '}}$. 
By complete analogy with the position space representation, $\tilde{g}_{\beta}(\vec{p}{\, '}-\vec{p})$ is interpreted as the quantum probability amplitude for the transition $\vec{p} \rightarrow \vec{p}{\, '}$ in smeared momentum space. 

Since an observed value ``$\vec{r}{\, '}$'' cannot determine which point(s) underwent the transition $\vec{r} \rightarrow \vec{r}{\, '}$ in the smeared superposition of geometries, we must sum over all possibilities by integrating the joint probability distribution $|\Psi(\vec{r},\vec{r}{\, '})|^2 := |\psi(\vec{r})|^2 |g(\vec{r}{\, '}-\vec{r})|^2$ over ${\rm d}^3\vec{r}$, yielding:
\begin{eqnarray} \label{EQ_XPRIMEDENSITY}
\frac{{\rm d}^{3}P(\vec{r}{\, '} | \Psi)}{{\rm d}\vec{r}{\, '}^{3}} = \int |\Psi(\vec{r},\vec{r}{\, '})|^2 {\rm 3}^d\vec{r} = (|\psi|^2 * |g|^2)(\vec{r}{\, '}) \, .
\end{eqnarray}
Here, physical predictions are assumed to be those of the smeared-space theory, and~the canonical QM of the original (unprimed) degrees of freedom is only a convenient tool in our calculations. 
In this formulation of the model, only primed degrees of freedom represent measurable quantities, whereas unprimed degrees of freedom are physically inaccessible \cite{Lake:2018zeg}.

The variance of a convolution is equal to the sum of the variances of the individual functions, so that the probability distribution (\ref{EQ_XPRIMEDENSITY}) gives rise to a GUR which is not of the canonical Heisenberg type. 
It is straightforward to verify that the same statistics can be obtained from the generalised position-measurement operator $\hat{X}^{i}$, defined as:

\begin{eqnarray} \label{X_operator}
\hat{X}^{i} := \int x'^{i} \, {\rm d}^{3} \hat{\mathcal{P}}_{\vec{r}{\, '}} = \hat{\mathbb{1}} \otimes \hat{x}'^{i} \, ,
\end{eqnarray}

where ${\rm d}^{3}\hat{\mathcal{P}}_{\vec{r}{\, '}} := \hat{\mathbb{1}} \otimes \ket{\vec{r}{\, '}}\bra{\vec{r}{\, '}}{\rm d}^{3}\vec{r}{\, '}$.
We then have
\begin{eqnarray} \label{X_uncertainty}
(\Delta_\Psi X^{i})^2 &=& \braket{\Psi |(\hat{X}^{i})^{2}|\Psi} - \braket{\Psi|\hat{X}^{i}|\Psi}^2
\nonumber\\
&=& (\Delta_\psi x'^{i})^2 + (\Delta_gx'^i)^2 \, .
\end{eqnarray}

Analogous reasoning in the momentum space representation gives
\begin{eqnarray} \label{EQ_PPRIMEDENSITY}
\frac{{\rm d}^{3}P(\vec{p}{\, '} | \tilde{\Psi})}{{\rm d}\vec{p}{\, '}^{3}} = \int |\tilde{\Psi}(\vec{p},\vec{p}{\, '})|^2 {\rm d}^{3}\vec{p} = (|\tilde{\psi}_{\hbar}|^2 * |\tilde{g}_{\beta}|^2)(\vec{p}{\, '}) \, ,
\end{eqnarray}
and 
\begin{eqnarray} \label{P_operator}
\hat{P}_{j} := \int p'_{j} \, {\rm d}^{3}\hat{\mathcal{P}}_{\vec{p}{\, '}} \, ,
\end{eqnarray}
where ${\rm d}^{3}\hat{\mathcal{P}}_{\vec{p}{\, '}} = \left(\int \ket{\vec{p}{\, '} \, \vec{p}{\, '}}\bra{\vec{p}{\, '} \, \vec{p}{\, '}} {\rm d}^{3}\vec{p}\right){\rm d}^{3}\vec{p}{\, '}$. 
It follows that
\begin{eqnarray} \label{P_uncertainty}
(\Delta_\Psi P_{j})^2 &=& \braket{\Psi |(\hat{P}_{i})^{2}|\Psi} - \braket{\Psi|\hat{P}_{j}|\Psi}^2
\nonumber\\
&=& (\Delta_\psi p'_{j})^2 + (\Delta_g p'_j)^2 \, .
\end{eqnarray}

Note that the HUP, expressed here in terms of primed variables
\begin{eqnarray} \label{HUP}
\Delta_{\psi} x'^{i} \Delta_{\psi} p'_{j} \geq \frac{\hbar}{2} \delta^{i}{}_{j} \, , 
\end{eqnarray}
(recall that the unprimed degrees of freedom are physically inaccessible) and the analogous relation
\begin{eqnarray} \label{Beta_UP}
\Delta_{g} x'^{i} \Delta_{g} p'_{j} \geq \frac{\beta}{2} \delta^{i}{}_{j} \, , 
\end{eqnarray}
both hold, independently of Equations (\ref{X_uncertainty}) and (\ref{P_uncertainty}). 
We denote the position and momentum uncertainties by $\Delta_{g}x'^{i} = \sigma_{g}^{i}$ and $\Delta_{g}p'_{j} = \tilde\sigma_{gj}$, respectively, when $|g|^2$ is chosen to be a Gaussian function. 
This saturates the inequality (\ref{Beta_UP}), yielding the definition of the Fourier transform scale $\beta$:
\begin{eqnarray} \label{beta}
\beta := (2/3)\sigma_{g}^{i}\tilde\sigma_{gi} \, . 
\end{eqnarray}

The HUP contains the essence of wave-particle duality, which could also be called wave-{\it point}-particle duality and is a fundamental consequence of the de Broglie relation $\vec{p} = \hbar \vec{k}$. 
This, in turn, is equivalent to the relation (\ref{dB-1}), which holds for particles propagating on a fixed (classical) Euclidean background.  
By contrast, Equation (\ref{Beta_UP}) represents the uncertainty relation for spatial ``points'' (not point-particles ``in'' space). 
This follows directly from Equation (\ref{mod_dB}), which is equivalent to the modified de Broglie relation:
\begin{eqnarray} \label{mod_dB*}
\vec{p}{\, '} = \hbar\vec{k} + \beta(\vec{k}'-\vec{k}) \, .
\end{eqnarray}
The new relation holds for particles propagating in the smeared-space background and the non-canonical term may be interpreted, heuristically, as an additional momentum ``kick'' induced by quantum fluctuations of the geometry.

Combining Equations (\ref{X_uncertainty}), (\ref{P_uncertainty}) and (\ref{HUP})--(\ref{Beta_UP}), gives
\begin{eqnarray} \label{GUR_X}
(\Delta_\Psi X^{i})^2 \, (\Delta_\Psi P_{j})^2 
&\ge&  \left(\frac{\hbar}{2}\right)^2(\delta^{i}{}_{j})^2 + (\Delta_\psi x'^{i})^2 (\Delta_{g} p'_{j})^2 
\nonumber\\
&+& (\Delta_{g}x'^{i})^2\frac{(\hbar/2)^2}{(\Delta_\psi x'^{j})^2} + (\Delta_{g}x'^{i})^2 (\Delta_{g} p'_{j})^2  \, ,
\end{eqnarray}
plus an analogous relation containing only $(\Delta_{\psi}p'_{j})^2$.
Optimising the right-hand side of (\ref{GUR_X}) with respect to $\Delta_\psi x'^{i}$, and~its counterpart with respect to $\Delta_\psi p'_{j}$, yields
\begin{equation} \label{EQ_CAN_DX_OPT}
(\Delta_\psi x'^{i})_{\mathrm{opt}} := \sqrt{\frac{\hbar}{2} \frac{\Delta_{g} x'^{i}}{\Delta_{g} p'_{i}}} \, , \quad (\Delta_\psi p'_{j})_{\mathrm{opt}} := \sqrt{\frac{\hbar}{2} \frac{\Delta_{g} p'_{j}}{\Delta_{g} x'^{j}}} \, , 
\end{equation}
so that
\begin{eqnarray} \label{DXDP_opt}
\Delta_\Psi X^{i} \, \Delta_\Psi P_{j} & \ge & \frac{(\hbar + \beta)}{2} \, \delta^{i}{}_{j} \, .
\end{eqnarray}
The same result is readily obtained by noting that the commutator of the generalised position and momentum observables is
\begin{equation} \label{[X,P]}
[\hat{X}^{i},\hat{P}_{j}] = i(\hbar + \beta)\delta^{i}{}_{j} \, {\bf\hat{\mathbb{I}}} \, ,
\end{equation}
where ${\bf\hat{\mathbb{I}}} = \hat{\mathbb{1}} \otimes \hat{\mathbb{1}}$ is the identity matrix on the tensor product space and $\hat{\mathbb{1}}$ is the identity matrix on the Hilbert space of canonical three-dimensional QM.
Equation (\ref{DXDP_opt}) then follows directly from the Schr{\" o}dinger--Robertson relation \cite{Ish95}. 
Thus, the inequalities in all three uncertainty relations, (\ref{HUP})--(\ref{Beta_UP}) and (\ref{GUR_X}), are saturated when $|g|^2$ is chosen to be a Gaussian, for which we denote $\Delta_{g}x'^{i} = \sigma_{g}^{i}$ and $\Delta_{g}p'_{j} = \tilde\sigma_{gj}$, and~when $|\psi|^2$ is chosen to be a Gaussian with $\Delta_\psi x'^{i} = (\Delta_\psi x'^{i})_{\rm opt}(\sigma_{g}^{i},\tilde\sigma_{gi})$, $\Delta_\psi p'_{j} = (\Delta_\psi p'_{j})_{\rm opt}(\sigma_{g}^{j},\tilde\sigma_{gj})$ (\ref{EQ_CAN_DX_OPT}). 
Importantly, the smeared-space model gives rise to minimum length and momentum uncertainties in the presence of commuting coordinates, i.e., 
\begin{equation} \label{XX_PP_commutators}
[\hat{X}^{i},\hat{X}^{j}] = 0 \, , \quad [\hat{P}_{i},\hat{P}_{j}] = 0 \, .
\end{equation}

Next, we note that setting 
\begin{equation} \label{sigma_g}
\sigma_g^{i} := \sqrt{2}l_{\rm Pl} \, , \quad \tilde{\sigma}_{gi} := \frac{1}{2}m_{\rm dS}c \, ,
\end{equation} 
where $l_{\rm Pl} := \sqrt{\hbar G/c^3}$ is the Planck length and $m_{\rm dS}c := \hbar/l_{\rm dS} := \hbar\sqrt{\Lambda/3} \simeq 10^{-66}$ g is the de Sitter mass, yields the required value of $\beta$ (\ref{beta_mag}). 
Using $l_{\rm dS} := \hbar/ m_{\rm dS}$ and $m_{\rm Pl} := \hbar/ l_{\rm Pl}$, we then have
\begin{eqnarray} \label{}
(\Delta_\psi x')_{\rm opt} = l_{\Lambda} \, , \quad (\Delta_\psi p')_{\rm opt} = \frac{1}{2}m_{\Lambda}c \, , 
\end{eqnarray}
where $l_{\Lambda} := 2^{1/4}\sqrt{l_{\rm Pl}l_{\rm dS}} \simeq 0.1 \ {\rm mm}$ and $m_{\Lambda} := 2^{-1/4}\sqrt{m_{\rm Pl}m_{\rm dS}} \simeq 10^{-3} \ {\rm eV}$. 
This gives rise to a minimum energy density of order: 
\begin{eqnarray} \label{min_energy_density}
\mathcal{E}_\psi \simeq \frac{3}{4\pi}\frac{(\Delta_\psi p')_{\rm opt}\ c}{(\Delta_\psi x')^3_{\rm opt}} \simeq \rho_{\Lambda}c^2 = \frac{\Lambda c^4}{8\pi G} \, , 
\end{eqnarray}
as required by current cosmological data \cite{Hobson:2006se}. 

In addition, using these values, Equations (\ref{X_uncertainty}), (\ref{P_uncertainty}) and (\ref{GUR_X}) may be Taylor expanded to first order to yield the GUP, EUP and EGUP, respectively, expressed in terms of $\Delta_{\psi} x'^{i}$ and $\Delta_{\psi} p'_{j}$. 
As discussed in the Introduction, the GUP implements a minimum length scale of the order of the Planck length $\sim l_{\rm Pl}$, but no minimum momentum scale, whereas the EUP implements a minimum momentum of the order of the de Sitter momentum $\sim m_{\rm dS}c$, but no minimum length. 
The EGUP therefore accounts for the effects of both minimum length and minimum momentum scales in nature. 
However, in the smeared-space model, neither $\Delta_{\psi} x'^{i}$ nor $\Delta_{\psi} p'_{j}$ are directly measurable, and~only $\Delta_{\Psi} X^{i}$ nor $\Delta_{\Psi} P_{j}$ are physical. 
It is therefore useful to express the smeared-space GUR directly in terms of these quantities. 
Thus, directly combining Equations (\ref{X_uncertainty}), (\ref{P_uncertainty}) and (\ref{HUP}), we obtain 
\begin{eqnarray} \label{smeared-spaceEGUP-1}
(\Delta_{\Psi} X^{i})^2 (\Delta_{\Psi} P_{j})^2 &\geq& \frac{\hbar^2}{4}(\delta^{i}{}_{j})^2 + (\sigma_g^i)^2(\Delta_{\Psi} P_{j})^2 
+ (\Delta_{\Psi} X^{i})^2(\tilde{\sigma}_{gj})^2 - (\sigma_g^i)^2(\tilde{\sigma}_{gj})^2 \, .
\end{eqnarray}
Setting $i=j$, we may ignore dimensional indices. 
Substituting for $\sigma_g$ and $\tilde{\sigma}_{g}$ from Equation (\ref{sigma_g}), taking the square root and expanding to first order, and~ignoring the subdominant term of order $\sim l_{\rm Pl}m_{\rm dS}c$ then yields the EGUP,
\begin{eqnarray} \label{smeared-spaceEGUP-2}
\Delta_{\Psi} X \Delta_{\Psi} P \gtrsim \frac{\hbar}{2} + \tilde{\alpha}(\Delta_{\Psi} P)^2 + \tilde{\eta}(\Delta_{\Psi} X)^2 \, ,
\end{eqnarray}
where 
\begin{eqnarray} \label{smeared-spaceEGUP-3}
\tilde{\alpha} = \frac{2G}{c^3} \, , \quad \tilde{\eta} = \hbar\frac{\Lambda}{12} \, .
\end{eqnarray}

Finally, we note that the smeared-space model has important implications for the description of measurement in quantum mechanics. 
We now illustrate these by considering a generalised position measurement in detail. 
Applying the generalised position operator $\hat{\vec{R}} := \hat{X}^{i}\underline{\bold{e}}_{i}(0)$ to an arbitrary pre-measurement state $\ket{\Psi}$ returns a random measured value, $\vec{r}{\, '}$, and~projects the state in the fixed background subspace of the tensor product onto: 
\begin{eqnarray} \label{X_measurement}
\ket{\psi_{\vec{r}{\, '}}} := \frac{1}{(|g|^2*|\psi|^2)(\vec{r}{\, '})}\int g(\vec{r}{\, '}-\vec{r}) \psi(\vec{r}) \ket{\vec{r}} {\rm d}^3\vec{r} \, , 
\end{eqnarray}
with probability $(|\psi|^2*|g|^2)(\vec{r}{\, '})$ \cite{Lake:2018zeg}. 
The total state is then $\ket{\psi_{\vec{r}{\, '}}}\otimes \ket{\vec{r}{\, '}}$, which is non-normalisable and therefore unphysical. 
This is analogous to the action of the canonical position measurement operator on $\ket{\psi}$, which projects onto the unphysical state $\ket{\vec{r}}$ with probability $|\psi|^2(\vec{r})$. 
However, in the smeared-space formalism, we must reapply the map (\ref{smear_map}) to complete the description of the measurement process. 
Thus, although generalised position measurements, represented by the application of the map (\ref{smear_map}) to the state (\ref{X_measurement}), yield precise measurement values, the post-measurement states are always physical, with well defined norms. 
Their position uncertainties, which may be determined by performing multiple measurements on ensembles of identically prepared systems, never fall below the fundamental smearing scale $\sigma_g \sim l_{\rm Pl}$. 
Analogous considerations hold for generalised momentum measurements, with the corresponding minimum value $\tilde{\sigma}_g \sim m_{\rm dS}c$. 

In this section, we have presented only a brief overview of the smeared-space formalism. 
The interested reader is referred to \cite{Lake:2018zeg} for further details.

\section{Angular Momentum in Smeared-Space QM} \label{Sec.4}

\subsection{A Simple Proposal} \label{Sec.4.1}

Clearly, the simplest way to construct a model of angular momentum for particles propagating on the smeared-space background is by defining the map:
\begin{eqnarray} \label{Map-1}
\vec{r} \mapsto \hat{\vec{R}} \, , \quad \vec{p} \mapsto \hat{\vec{P}} \, , 
\end{eqnarray} 
where 
\begin{eqnarray} \label{Map-2}
\hat{\vec{R}} := \hat{X}^i\underline{\bold{e}}_i(0) \, , \quad \hat{\vec{P}} := \hat{P}_j\underline{\bold{e}}^j(0) \, ,
\end{eqnarray}
and $\hat{X}^i$, $\hat{P}_j$ are the generalised operators given by Equations (\ref{X_operator}) and (\ref{P_operator}), respectively. 
We may then define the smeared-space angular momentum operator as:
\begin{eqnarray} \label{L}
\hat{\vec{L}} = \hat{\vec{R}} \times \hat{\vec{P}} \, . 
\end{eqnarray}

However, considering the arguments presented in Section \ref{Sec.2} (see also Appendices \ref{Appendix.A} and \ref{Appendix.B}) the validity of Equation (\ref{Map-2}) relies on the stricter interpretation of the smearing function, $g(\vec{r}{\, '}-\vec{r})$, suggested therein. 
Hence, from here on, we interpret $g(\vec{r}{\, '}-\vec{r})$ as the probability amplitude for the transition $\vec{r} \leftrightarrow \vec{r}{\, '}$. 
Such pair-wise exchanges do not affect the curvature of the background geometry, so that we may continue to use Cartesian tangent vectors in our analysis.

The Cartesian components of $\hat{\vec{L}} := \hat{L}_{i}\underline{\bold{e}}^i(0)$ are then given by
\begin{eqnarray} \label{L_i}
\hat{L}_{i} = (\hat{\vec{R}} \times \hat{\vec{P}})_{i} = \epsilon_{ij}{^{k}}\hat{X}^{j}\hat{P}_{k} \, ,
\end{eqnarray}
and may be obtained directly from the standard inner product, $\hat{L}_{i} = \braket{\underline{\bold{e}}_{i}(0),\hat{\vec{L}}} = \hat{L}_{j}\braket{\underline{\bold{e}}_{i}(0),\underline{\bold{e}}^{j}(0)} = \hat{L}_{j}\delta^{j}{}_{i}$. 
From Equations (\ref{[X,P]}), (\ref{XX_PP_commutators}) and (\ref{L_i}), it also follows that:
\begin{eqnarray} \label{commutators_LX_LP}
[\hat{L}_{i},\hat{X}^{k}] = i(\hbar + \beta) \, \epsilon_{ij}{}^{k} \hat{X}^{j} \, , \quad [\hat{L}_{i},\hat{P}_{j}] = i(\hbar + \beta) \, \epsilon_{ij}{}^{k} \hat{P}_{k}
\end{eqnarray}
and
\begin{eqnarray} \label{LL_commutator}
[\hat{L}_{i},\hat{L}_{j}] = i(\hbar +\beta) \epsilon_{ij}{}^{k}\hat{L}_{k} \, ,  
\end{eqnarray}
\begin{eqnarray} \label{L^2L_commutator}
[\hat{L}^2,\hat{L}_{i}] = 0 \, .
\end{eqnarray}
Hence,
\begin{eqnarray} \label{L_vec_commutator}
[\hat{L}_{\vec{a}},\hat{L}_{\vec{b}}] = i(\hbar +\beta) \sin\theta \hat{L}_{\underline{\bold{n}}} \, ,
\end{eqnarray}
where $\hat{L}_{\vec{a}}$ denotes the projection of $\hat{\vec{L}}$ onto an arbitrary unit vector $\vec{a}$ and $\underline{\bold{n}} = \vec{a} \times \vec{b}$.

By the Schr{\" o}dinger--Robertson relation, Equations (\ref{LL_commutator}) and (\ref{L_vec_commutator}) give rise to the uncertainty relations
\begin{eqnarray} \label{LL_UR}
\Delta_{\Psi}L_{i}\Delta_{\Psi}L_{j} \geq \frac{(\hbar +\beta)}{2} |\epsilon_{ij}{^{k}}\braket{\hat{L}_{k}}_{\Psi}| \, ,  
\end{eqnarray}
\begin{eqnarray} \label{L_vec_UR}
\Delta_{\Psi}L_{\vec{a}}\Delta_{\Psi}L_{\vec{b}} \geq \frac{(\hbar +\beta)}{2} |\sin\theta \braket{\hat{L}_{\underline{\bold{n}}}}_{\Psi}| \, ,
\end{eqnarray}
respectively.
These are completely analogous to Equations (\ref{ll_commutator}) and (\ref{l^2l_commutator}) but with the rescaling $\hbar \rightarrow \hbar + \beta$. 
As in canonical QM, Equation (\ref{L_vec_UR}) is the more general relation, and~the uncertainty relation for Cartesian components is recovered by taking $\theta = m(\pi/2)$ ($m \in \mathbb{Z}$). 
Similarly, Equations (\ref{commutators_LX_LP}) give rise to
\begin{eqnarray} \label{UR_LX_LP}
\Delta_{\Psi}L_{i}\Delta_{\Psi}X^{k} \geq \frac{(\hbar + \beta)}{2} \, |\epsilon_{ij}{}^{k} \braket{\hat{X}^{j}}_{\Psi}| \, , \quad
\Delta_{\Psi}L_{i}\Delta_{\Psi}P_{j} \geq \frac{(\hbar + \beta)}{2}\, |\epsilon_{ij}{}^{k} \braket{\hat{P}_{k}}_{\psi}| \, ,
\end{eqnarray}
which are analogous to Equations (\ref{UR_lx_lp}).

However, since Equations (\ref{commutators_LX_LP})--(\ref{UR_LX_LP}) are of almost canonical form, it is not immediately clear how (or why) the smeared-space model should generate GURs for angular momentum. 
Nonetheless, in the remainder of this section, we will show that such GURs {\it are} generated. 
The key point is that, although the model implies only a simple rescaling of the canonical Schr{\" o}dinger--Robertson bound, for any pair of operators, it nonetheless generates GURs of the form $\Delta_{\Psi}X^{i}\Delta_{\Psi}P_{j} \geq \dots \geq (\hbar + \beta)/2 \, . \, \delta^{i}{}_{j}$, $\Delta_{\Psi}L_{i}\Delta_{\Psi}L_{j} \geq \dots \geq (\hbar + \beta)/2 \, . \, |\epsilon_{ij}{^{k}}\braket{\hat{L}_{k}}_{\Psi}|$, etc. 
The dots in the middle of each of these expressions represent a sum of terms which is generically larger than the Schr{\" o}dinger--Robertson bound on the far right-hand side. 
The resulting hierarchy of inequalities is saturated only under specific conditions, in which the terms in the middle are optimised with respect to the relevant variables.

Thus, in order to gain deeper insight into the behaviour of angular momentum in the smeared-space model, we must investigate the origins of the relations (\ref{commutators_LX_LP})--(\ref{UR_LX_LP}) in more detail. 
Below, we show explicitly that, despite their canonical form (except for the rescaling $\hbar \rightarrow \hbar + \beta$), Equations (\ref{LL_UR}), (\ref{L_vec_UR}) and (\ref{UR_LX_LP}) are compatible with GURs for angular momentum. 
In this sense, they are analogous to Equation (\ref{DXDP_opt}), which, despite its canonical form (except for $\hbar \rightarrow \hbar + \beta$), is compatible with the GURs (\ref{GUR_X}) and (\ref{smeared-spaceEGUP-1})--(\ref{smeared-spaceEGUP-2}). 

\subsection{Useful Alternative Formalism} \label{Sec.4.2}

To investigate the structure of the generalised commutator $[\hat{L}_{i},\hat{L}_{j}]$ in more detail it is useful to first rewrite the generalised position and momentum operators, $\hat{X}^{i}$ and $\hat{P}_{j}$, as well as the smeared-state $\ket{\Psi}$, in a unitarily equivalent form. We begin by constructing the operator
\begin{eqnarray} \label{U_beta}
\hat{U}_{\beta} &:=& \exp\left[\frac{i}{\beta}(\hat{\mathbb{1}} \otimes \hat{\vec{p}}{\, '}).(\hat{\vec{r}} \otimes \hat{\mathbb{1}}) \right] \, , 
\end{eqnarray}
whose action on the smeared-space basis is
\begin{eqnarray} \label{U_beta_action_pos}
\hat{U}_{\beta} \ket{\vec{r},\vec{r}{\, '}} = \ket{\vec{r},\vec{r}{\, '} - \vec{r}} \, ,
\end{eqnarray}
\begin{eqnarray} \label{U_beta_action_mom}
\hat{U}_{\beta} \ket{\vec{p} \,\vec{p}{\, '}} = \ket{\vec{p},\vec{p}{\, '} - \vec{p}} \, .
\end{eqnarray}

Here, we again assume that, while $\hbar$ sets the quantisation scale for the degrees of freedom in the first subspace of the tensor product, $\beta$ sets the quantisation scale for the degrees of freedom in the second subspace. 
Hence, $\beta^{-1}(\hat{\mathbb{1}} \otimes \hat{\vec{p}}{\, '})$ generates translations on the second vector of the basis $\ket{\vec{r},\vec{r}{\, '}}$, just as $\hbar^{-1}(\hat{\vec{p}} \otimes \hat{\mathbb{1}})$ generates translations on the first. 
This accounts for Equation (\ref{U_beta_action_pos}). 
Equation (\ref{U_beta_action_mom}) then follows by combining Equations (\ref{U_beta})--(\ref{U_beta_action_pos}) with (\ref{mod_dB}), again assuming that $\beta$ sets the Fourier transform scale for kets in the second subspace (as required for consistency).

Together, these considerations imply
\begin{eqnarray} \label{non-canon*}
\braket{\vec{r},\vec{r}{\, '} | \vec{p} \,\vec{p}{\, '}} = \braket{\vec{r} | \vec{p}}_1\braket{\vec{r}{\, '} - \vec{r} | \vec{p}{\, '} - \vec{p}}_2 \, , 
\end{eqnarray}
where
\begin{eqnarray} \label{non-canon-3}
\braket{\vec{r} | \vec{p}}_1 = \frac{1}{\sqrt{2\pi\hbar}}  e^{\frac{i}{\hbar}\vec{p}.\vec{r}} 
\end{eqnarray}
(as in canonical QM) and
\begin{eqnarray} \label{non-canon-4}
\braket{\vec{r}{\, '} - \vec{r} | \vec{p}{\, '} - \vec{p}}_2 = \frac{1}{\sqrt{2\pi\beta}} e^{\frac{i}{\beta}(\vec{p}{\, '}-\vec{p}).(\vec{r}{\, '}-\vec{r})} \, .
\end{eqnarray}
In Equations (\ref{non-canon*})--(\ref{non-canon-4}), we use the subscripts 1 and 2 to indicate which subspace of the tensor product state the brakets belong to. 
This is to avoid confusion since, in these expressions, the degrees of freedom in each subspace are no longer labelled exclusively by primed or unprimed variables, as they were previously. 
Nonetheless, they remain consistent with our convention that $\hbar$ sets the quantisation scale for degrees of freedom in the first subspace of the tensor product, while $\beta$ sets the quantisation scale for degrees of freedom in the second. 
We repeat that the former are associated with canonical quantum matter whereas the latter are associated with the quantum state of the background geometry. 

Using these results, we map the smeared-space operators $\hat{X}^{i}$ and $\hat{P}_{j}$, and~the smeared-state $\ket{\Psi}$, according to
\begin{eqnarray} \label{X_unitary_equiv}
\hat{X}^{i} &\mapsto& \hat{U}_{\beta}^{\dagger} \hat{X}^{i} \hat{U}_{\beta}  
\nonumber\\
&=& \int\int x'^{i} \ket{\vec{r}}\bra{\vec{r}} \otimes \ket{\vec{r}{\, '} - \vec{r}}\bra{\vec{r}{\, '} - \vec{r}}{\rm d}^3\vec{r} {\rm d}^3\vec{r}{\, '} 
\end{eqnarray}
\begin{eqnarray} \label{P_unitary_equiv}
\hat{P}_{j} &\mapsto& \hat{U}_{\beta}^{\dagger} \hat{P}_{j} \hat{U}_{\beta}  
\nonumber\\
&=& \int\int p'_{j} \ket{\vec{p}}\bra{\vec{p}} \otimes \ket{\vec{p}{\, '} - \vec{p}}\bra{\vec{p}{\, '} - \vec{p}}{\rm d}^3\vec{p} {\rm d}^3\vec{p}{\, '} 
\end{eqnarray}
and
\begin{eqnarray} \label{Psi_unitary_equiv}
\ket{\Psi} &\mapsto& \hat{U}_{\beta}\ket{\Psi}
\nonumber\\
&=& \int\int g(\vec{r}{\, '} - \vec{r})\psi(\vec{r}) \ket{\vec{r},\vec{r}{\, '}-\vec{r}} {\rm d}^3\vec{r} {\rm d}^3\vec{r}{\, '}
\nonumber\\
&=& \int\int \tilde{g}_{\beta}(\vec{p}{\, '} - \vec{p})\tilde{\psi}_{\hbar}(\vec{p}) \ket{\vec{p},\vec{p}{\, '}-\vec{p}} {\rm d}^3\vec{p} {\rm d}^3\vec{p}{\, '} 
\nonumber\\
&=:& \ket{\psi} \otimes \ket{g} \, . 
\end{eqnarray}
Note that Equation (\ref{Psi_unitary_equiv}) implicitly defines the state $\ket{g}$, which is distinct from the state $\ket{g_{\vec{r}}}$ defined in Equation (\ref{|g_r>}). 
Physically, $\ket{g_{\vec{r}}}$ represents the quantum state associated with the smeared ``point'' $\vec{r}$, whereas $\ket{g}$ is the quantum state associated with whole background space. 
From here on, we use $\hat{X}^{i}$, $\hat{P}_{j}$ and $\ket{\Psi}$ to refer to the unitarily equivalent forms of the generalised position and momentum operators (\ref{X_unitary_equiv}) and (\ref{P_unitary_equiv}), and~smeared-state (\ref{Psi_unitary_equiv}), respectively, unless stated otherwise. 

Next, we split each of the generalised operators (\ref{X_unitary_equiv}) and (\ref{P_unitary_equiv}) into the sum of two terms:
\begin{eqnarray} \label{ops_split_Q}
\hat{X}^{i} = \hat{Q}^{i} + \hat{Q}'^{i} = (\hat{q}^{i} \otimes \hat{\mathbb{1}}) + (\hat{\mathbb{1}} \otimes \hat{q}'^{i}) \, , 
\end{eqnarray}
\begin{eqnarray} \label{ops_split_Pi}
\hat{P}_{j} = \hat{\Pi}_{j} + \hat{\Pi}'_{j} = (\hat{\pi}_{j} \otimes \hat{\mathbb{1}}) + (\hat{\mathbb{1}} \otimes \hat{\pi}'_{j}) \, , 
\end{eqnarray}
where
\begin{eqnarray} \label{QQ'}
\hat{Q}^{i} &:=& \int\int x^{i} \ket{\vec{r}}\bra{\vec{r}} \otimes \ket{\vec{r}{\, '} - \vec{r}}\bra{\vec{r}{\, '} - \vec{r}}{\rm d}^3\vec{r} {\rm d}^3\vec{r}{\, '} 
\nonumber\\
&:=& (\hat{q}^{i} \otimes \hat{\mathbb{1}}) \, , 
\nonumber\\
\hat{Q}'^{i} &:=& \int\int (x'^{i} -x^{i}) \ket{\vec{r}}\bra{\vec{r}} \otimes \ket{\vec{r}{\, '} - \vec{r}}\bra{\vec{r}{\, '} - \vec{r}}{\rm d}^3\vec{r} {\rm d}^3\vec{r}{\, '} 
\nonumber\\
&:=& (\hat{\mathbb{1}} \otimes \hat{q}'^{i}) \, ,
\end{eqnarray}
and 
\begin{eqnarray} \label{PiPi'}
\hat{\Pi}_{j} &:=& \int\int p_{j} \ket{\vec{p}}\bra{\vec{p}} \otimes \ket{\vec{p}{\, '} - \vec{p}}\bra{\vec{p}{\, '} - \vec{p}}{\rm d}^3\vec{p} {\rm d}^3\vec{p}{\, '} 
\nonumber\\
&:=& (\hat{\pi}_{j} \otimes \hat{\mathbb{1}}) \, , 
\nonumber\\
\hat{\Pi}'_{j} &:=& \int\int (p'_{j} -p_{j}) \ket{\vec{p}}\bra{\vec{p}} \otimes \ket{\vec{p}{\, '} - \vec{p}}\bra{\vec{p}{\, '} - \vec{p}}{\rm d}^3\vec{p} {\rm d}^3\vec{p}{\, '} 
\nonumber\\
&:=& (\hat{\mathbb{1}} \otimes \hat{\pi}'_{j}) \, .
\end{eqnarray}
In other words, we define the new classical variables
\begin{eqnarray} \label{new_variables-1}
\vec{Q} := \vec{r}{\, '} \, , \quad \vec{q} := \vec{r} \, , \quad \vec{q}{\, '} := (\vec{r}{\, '} - \vec{r}) \, , 
\end{eqnarray}
\begin{eqnarray} \label{new_variables-2}
\vec{\Pi} := \vec{p}{\, '} \, , \quad \vec{\pi} := \vec{p} \, , \quad \vec{\pi}{\, '} := (\vec{p}{\, '} - \vec{p}) \, , 
\end{eqnarray}
and construct their quantum operator counterparts. 

Note that, in this formulation of the smeared-space model, measurable quantities are no longer expressed in terms of primed variables only.  
That is, neither $q^{i}$ nor $q'^{i}$ are measurable, individually, and~only their sum $q^{i} + q'^{i} = x'^{i}$ is physical. 
Similarly, neither $\pi_{j}$ nor $\pi'_{j}$ is directly measurable, only $\pi_{j} + \pi'_{j} = p'_{j}$. 
This has important physical consequences. 

In the first formulation of the model \cite{Lake:2018zeg}, summarised in Section \ref{Sec.3}, the wave functions corresponding to matter and geometry are entangled, as hypothesised in \cite{Kay:2018mxr}. 
However, in the unitarily equivalent formulation, presented here, they are not. 
Nonetheless, {\it physical} measurements correspond to operators that act on both subsytems of the tensor product state $\ket{\Psi}$, regardless of our choice of basis. 
Furthermore, since the basis transformation (\ref{U_beta_action_pos}) is a unitary operation, the effects of geometry-matter entanglement (in the first formulation) cannot be ``undone'' by this change. 
In other words, although the entanglement of states is basis-dependent, and~therefore not fundamental, predictions for the results of physical measurements arise from the combination of both states and operators. 
These predictions are basis-independent, as required. 

From Equations (\ref{QQ'}) and (\ref{PiPi'}) it is straightforward to show that 
\begin{eqnarray} \label{[Q,Pi]}
[\hat{Q}^{i},\hat{\Pi}_{j}] = i\hbar \delta^{i}{}_{j} \, {\bf\hat{\mathbb{I}}} \, , \quad [\hat{Q}'^{i},\hat{\Pi}'_{j}]  = i\beta \delta^{i}{}_{j} \, {\bf\hat{\mathbb{I}}} \, .
\end{eqnarray}
and
\begin{eqnarray} \label{mixed_comm-1}
[\hat{Q}^{i},\hat{\Pi}'_{j}] = [\hat{Q}'^{i},\hat{\Pi}_{j}]  = 0 \, .
\end{eqnarray}
Together, Equations (\ref{[Q,Pi]}) and (\ref{mixed_comm-1}) recover Equation (\ref{[X,P]}). 
The remaining commutation relations are
\begin{eqnarray} \label{QQ_commutators}
[\hat{Q}^{i},\hat{Q}^{j}] = [\hat{Q}'^{i},\hat{Q}'^{j}] = 0 \, ,
\end{eqnarray}
\begin{eqnarray} \label{PiPi_commutators}
\quad [\hat{\Pi}_{i}, \hat{\Pi}_{j}] = [\hat{\Pi}'_{i}, \hat{\Pi}'_{j}] = 0 \, ,
\end{eqnarray}
and
\begin{eqnarray} \label{remaining_commutators}
[\hat{Q}^{i},\hat{Q}'^{j}] = 0 \, , \quad [\hat{\Pi}_{i}, \hat{\Pi}'_{j}] = 0 \, . 
\end{eqnarray}

We then have
\begin{eqnarray} \label{comp-1}
(\Delta_{\Psi}X^{i})^2 &=& (\Delta_{\Psi}Q^{i})^2 + (\Delta_{\Psi}Q'^{i})^2 \, ,
\end{eqnarray}
and
\begin{eqnarray} \label{comp-2}
(\Delta_{\Psi}P_{j})^2 &=& (\Delta_{\Psi}\Pi_{j})^2 + (\Delta_{\Psi}\Pi'_{j})^2 \, , 
\end{eqnarray}
since ${\rm cov}_{\Psi}(\hat{Q}^{i},\hat{Q}'^{i}) = {\rm cov}_{\Psi}(\hat{Q}'^{i},\hat{Q}^{i}) = 0$ and 
${\rm cov}_{\Psi}(\hat{\Pi}_{j},\hat{\Pi}'_{j}) = {\rm cov}_{\Psi}(\hat{\Pi}'_{j},\hat{\Pi}_{j}) = 0$, where ${\rm cov}(X,Y) = \braket{XY} - \braket{X}\braket{Y}$ is the covariance of the random variables $X$ and $Y$. 
The operator pairs $\hat{Q}^{i},\hat{Q}'^{i}$ and $\hat{\Pi}_{j},\hat{\Pi}'_{j}$ are uncorrelated since they act on separate subspaces of the total state $\ket{\Psi}$. 
Comparison of Equations (\ref{comp-1}) and (\ref{comp-2}) with Equations (\ref{X_uncertainty}) and (\ref{P_uncertainty}), respectively, suggests
\begin{eqnarray} \label{}
\Delta_{\Psi}Q^{i} = \Delta_{\psi}x'^{i} \, , \quad \Delta_{\Psi}Q'^{i} = \Delta_{g}x'^{i} \, , 
\end{eqnarray}
\begin{eqnarray} \label{}
\Delta_{\Psi}\Pi_{j} = \Delta_{\psi}p'_{j} \, , \quad \Delta_{\Psi}\Pi'_{j} = \Delta_{g}p'_{j} \, ,
\end{eqnarray}
and it is straightforward to verify these equalities explicitly. 
Furthermore, in the position and momentum space representations of smeared-space wave mechanics, the new operators take the especially simple forms:
\begin{eqnarray} \label{QQ'_pos}
\hat{Q}^{i} = x^{i}= q^{i} \, , \quad \hat{Q}'^{i} = (x'^{i} - x^{i}) = q'^{i}\, , 
\end{eqnarray}
\begin{eqnarray} \label{PiPi'_pos}
\hat{\Pi}_{j} &=& -i\hbar \frac{\partial}{\partial x^{j}}\Bigg|_{(\vec{r}{\, '}-\vec{r}) = {\rm const.}} = -i\hbar \frac{\partial}{\partial q^{j}}\Bigg|_{\vec{q}{\, '}= {\rm const.}} \, , 
\nonumber\\
\hat{\Pi}'_{j} &=& -i\beta \frac{\partial}{\partial (x'^{j}-x^{j})}\Bigg|_{\vec{r} = {\rm const.}} = -i\beta \frac{\partial}{\partial q'^{j}}\Bigg|_{\vec{q} = {\rm const.}} \, , 
\end{eqnarray}
and 
\begin{eqnarray} \label{QQ'_mom}
\hat{Q}^{i} &=& i\hbar \frac{\partial}{\partial p_{i}}\Bigg|_{(\vec{p}{\, '}-\vec{p}) = {\rm const.}} = i\hbar \frac{\partial}{\partial \pi_{i}}\Bigg|_{\vec{\pi}{\, '}= {\rm const.}} \, , 
\nonumber\\
\hat{Q}'^{i} &=& i\beta \frac{\partial}{\partial (p'_{i}-p_{i})}\Bigg|_{\vec{p} = {\rm const.}} = i\beta \frac{\partial}{\partial \pi'_{i}}\Bigg|_{\vec{\pi} = {\rm const.}} \, , 
\end{eqnarray}
\begin{eqnarray} \label{PiPi'_mom}
\hat{\Pi}_{j} = p_{j} = \pi_{j} \, , \quad \hat{\Pi}'_{j} = (p'_{j} - p_{j}) = \pi'_{j}\, , 
\end{eqnarray}
respectively. 

We now see the origin of the rescaled commutation relation (\ref{[X,P]}) more clearly. 
The linear momentum of the matter sector generates one copy of the shift-isometry algebra, scaled by $\hbar$. 
This is the canonical Heisenberg algebra for a quantum point-particle. 
Simultaneously, the linear momentum of quantum ``points'' in the background generates an additional copy of the algebra, scaled by $\beta$. 
The two representations commute and, since each copy of the position-momentum commutator is proportional to the same constant, $\delta^{i}{}_{j}$, we are left with a single factor of $(\hbar + \beta)/2 \, . \, \delta^{i}{}_{j}$ on the right-hand side of Equation (\ref{[X,P]}).

This analysis clearly shows that the ultimate origin of both the rescaling $\hbar \rightarrow \hbar + \beta$ {\it and} the GURs (\ref{comp-1}) and (\ref{comp-2}) is the generalised algebra represented by Equations (\ref{[Q,Pi]})--(\ref{remaining_commutators}). 
In Section \ref{Sec.4.3}, we use the alternative formalism presented here to derive the corresponding algebra for smeared angular momentum operators. 
By analogy with our previous results we show that this generates a simple rescaling of the canonical ${\rm so}(3)$ Lie algebra together with GURs for angular momentum.

Before concluding this section, we stress that the unitary operator (\ref{U_beta}) is not a function of the {\it physical} position and momentum operators of the smeared-space model, 
$\hat{U}_{\beta} \neq \hat{U}_{\beta}(\hat{\vec{R}},\hat{\vec{P}})$. 
In terms of our new variables, (\ref{ops_split_Q}) and (\ref{ops_split_Pi}), it may be written as $\hat{U}_{\beta} = \exp\left[(i/\beta)\hat{\vec{\Pi}}'.\hat{\vec{Q}}\right]$, but neither $\hat{\vec{Q}}$ nor $\hat{\vec{\Pi}}'$, alone, represents the physical (i.e., measurable) position or momentum of the particle. 
Hence, although we may construct $\hat{U}_{\beta}$ mathematically, and~utilise it to simplify our calculations, it is doubtful that this represents a viable physical transformation of the system that could actually be carried out in the ``real'' smeared-space universe.

Nonetheless, it is interesting to note that, formally, $\hat{U}_{\beta}$ is analogous to the unitary operator that implements transitions between quantum reference frames (QRFs) defined in \cite{Giacomini:2017zju}. 
The primary difference is the presence of the parameter $\beta$, in place of $\hbar$, though this is to be expected since the formalism defined in \cite{Giacomini:2017zju} corresponds to canonical quantum systems on fixed backgrounds. 
Interpreted in this way, the original formulation of the smeared-space model, in which the momentum space basis is given by Equation (\ref{mod_dB}), represents the view of the composite state $\ket{\Psi}$ from a single QRF in which matter and geometry (i.e., $\psi$ and $g$) are entangled. 
Thus, if $\hat{U}_{\beta}$ represents a physical transformation of the system, the reformulation of the model presented here represents the view of the composite state of matter and geometry from a new QRF, in which this state is {\it separable}.
However, as mentioned above, it is doubtful that $\hat{U}_{\beta}$ represents a physically realisable transformation. 
Despite this, the formal similarity between the two models is suggestive of a more profound and genuine link, which could yield insights into models of quantum gravity. 
We aim to explore this possibility in a future work.

\subsection{Generalised Algebra and GURs} \label{Sec.4.3}

In terms of our new operators (\ref{ops_split_Q}) and (\ref{ops_split_Pi}), the components of the generalised angular momentum may be written as:
\begin{eqnarray} \label{L_sum}
\hat{L}_{i} = \hat{\mathcal{L}}_{i} + \hat{\mathcal{L}}'_{i} + \hat{\Lambda}_{i} + \hat{\Lambda}'_{i} \, , 
\end{eqnarray}
where
\begin{eqnarray} \label{L_terms}
\hat{\mathcal{L}}_{i} &:=& \epsilon_{ij}{}^{k} \hat{Q}^{j}\hat{\Pi}_{k} \, , \quad \hat{\mathcal{L}}'_{i} := \epsilon_{ij}{}^{k} \hat{Q}'^{j}\hat{\Pi}'_{k} \, ,
\nonumber\\
\hat{\Lambda}_{i} &:=& \epsilon_{ij}{}^{k} \hat{Q}^{j}\hat{\Pi}'_{k} \, , \quad \hat{\Lambda}'_{i} := \epsilon_{ij}{}^{k} \hat{Q}'^{j}\hat{\Pi}_{k} \, .
\end{eqnarray}
After straightforward (but tedious) algebraic manipulation, it may be verified that the individual subcomponents $\left\{\hat{\mathcal{L}}_{i},\hat{\mathcal{L}}'_{i},\hat{\Lambda}_{i},\hat{\Lambda}'_{i}\right\}$ of the generalised generators (\ref{L_sum}) obey the following algebra:
%
\begin{equation} \label{rearrange-L.1}
[\hat{\mathcal{L}}_{i},\hat{\mathcal{L}}_{j}] = i\hbar \, \epsilon_{ij}{}^{k} \hat{\mathcal{L}}_{k} \, , \quad [\hat{\mathcal{L}}'_{i},\hat{\mathcal{L}}'_{j}] = i\beta \, \epsilon_{ij}{}^{k} \hat{\mathcal{L}}'_{k} \, , 
\end{equation}
\begin{equation} \label{rearrange-L.2}
[\hat{\mathcal{L}}_{i},\hat{\mathcal{L}}'_{j}] = [\hat{\mathcal{L}}'_{i},\hat{\mathcal{L}}_{j}] = 0 \, , 
\end{equation}
\begin{equation} \label{rearrange-L.3}
[\hat{\mathcal{L}}_{i},\hat{\Lambda}_{j}] - [\hat{\mathcal{L}}_{j},\hat{\Lambda}_{i}] = i\hbar \, \epsilon_{ij}{}^{k} \hat{\Lambda}_{k} \, , 
\end{equation}
\begin{equation} \label{rearrange-L.4}
[\hat{\mathcal{L}}_{i},\hat{\Lambda}'_{j}] - [\hat{\mathcal{L}}_{j},\hat{\Lambda}'_{i}] = i\hbar \, \epsilon_{ij}{}^{k} \hat{\Lambda}'_{k} \, , 
\end{equation}
\begin{equation} \label{rearrange-L.5}
[\hat{\mathcal{L}}'_{i},\hat{\Lambda}_{j}] - [\hat{\mathcal{L}}'_{j},\hat{\Lambda}_{i}] = i\beta \, \epsilon_{ij}{}^{k} \hat{\Lambda}_{k} \, , 
\end{equation}
\begin{equation} \label{rearrange-L.6}
[\hat{\mathcal{L}}'_{i},\hat{\Lambda}'_{j}] - [\hat{\mathcal{L}}'_{j},\hat{\Lambda}'_{i}] = i\beta \, \epsilon_{ij}{}^{k} \hat{\Lambda}'_{k} \, , 
\end{equation}
\begin{equation} \label{rearrange-L.7}
[\hat{\Lambda}_{i},\hat{\Lambda}_{j}] = [\hat{\Lambda}'_{i},\hat{\Lambda}'_{j}] = 0 \, , 
\end{equation}
\begin{equation} \label{rearrange-L.8}
[\hat{\Lambda}_{i},\hat{\Lambda}'_{j}] - [\hat{\Lambda}_{j},\hat{\Lambda}'_{i}] = i\beta \, \epsilon_{ij}{}^{k}\hat{\mathcal{L}}_{k} + i\hbar \, \epsilon_{ij}{}^{m}\hat{\mathcal{L}}'_{m} \, .
\end{equation}
%
Note that summing the left-hand sides of Equations (\ref{rearrange-L.1})--(\ref{rearrange-L.8}) yields the generalised commutator $[\hat{L}_{i},\hat{L}_{j}]$ whereas summing the right-hand sides yields $i(\hbar + \beta)\, \epsilon_{ij}{}^{k} \hat{L}_{k}$, as required.

Equations (\ref{rearrange-L.1}) confirm that $\hat{\mathcal{L}}_{i}$ and $\hat{\mathcal{L}}'_{i}$ represent genuine angular momentum operators since the subsets $\left\{\hat{\mathcal{L}}_{i}\right\}_{i=1}^{3}$ and $\left\{\hat{\mathcal{L}}'_{i}\right\}_{i=1}^{3}$ satisfy the required algebras, that is, appropriately scaled representations of ${\rm so}(3)$. 
According to our previous interpretation of the tensor product state (\ref{Psi_unitary_equiv}) $\hat{\mathcal{L}}_{i}$ represents the angular momentum of the canonical quantum state vector $\ket{\psi}$ (quantised at the scale $\hbar$) whereas $\hat{\mathcal{L}}'_{i}$ represents the angular momentum associated with the quantum state of the background $\ket{g}$ (quantised at the scale $\beta$).
By contrast, Equations (\ref{rearrange-L.7}) and (\ref{rearrange-L.8}) show that $\hat{\Lambda}_{i}$ and $\hat{\Lambda}'_{i}$ do not represent components of angular momentum in their own right. 
These ``cross terms'' determine the effect, on the angular momentum of a canonical quantum particle, of its interaction with the smeared background. 

We also note that, since neither $\hat{\Lambda}_{i}$ nor $\hat{\Lambda}'_{i}$ commute with either $\hat{\mathcal{L}}_{i}$ or $\hat{\mathcal{L}}'_{i}$, it is impossible for a smeared state $\ket{\Psi}$ to be an eigenvector of all four subcomponents of $\hat{L}_{i}$ simultaneously. 
Nonetheless, Equation (\ref{L^2L_commutator}) demonstrates that the simultaneous eigenvectors of $\hat{L}^2$ and $\hat{L}_{i}$ form a valid basis of the infinite-dimensional Hilbert space $\mathcal{H} \otimes \mathcal{H}'$ ($\mathcal{H}' \cong \mathcal{H}$).
In other words, if both $\ket{\psi}$ and $\ket{g}$ represent angular momentum eigenstates (that is, if $\hat{\mathcal{L}}_{i}\ket{\Psi} = m\hbar\ket{\Psi}$ and $\hat{\mathcal{L}}'_{i}\ket{\Psi} = m'\beta\ket{\Psi}$ for $m,m' \in \mathbb{Z}$), then the total state $\ket{\Psi} = \ket{\psi} \otimes \ket{g}$ is {\it not} an eigenstate of $\hat{L}_{i}$. 
In this way, single-particle smeared-states differ starkly from unentangled bipartite states in canonical QM: $\ket{\psi_{\rm tot}} = \ket{\psi_{1}} \otimes \ket{\psi_{2}}$.
 
Alternatively, we may write the generalised operator $\hat{L}_i$ as:
\begin{eqnarray} \label{L_sum*}
\hat{L}_{i} = \hat{\mathcal{L}}_{i} + \hat{\mathcal{L}}'_{i} + \hat{\mathbb{L}}_{i} \, , 
\end{eqnarray}
where
\begin{eqnarray} \label{mathb{L}_{i}}
\hat{\mathbb{L}}_{i} := \hat{\Lambda}_{i} + \hat{\Lambda}'_{i} \, .
\end{eqnarray}
The new subcomponents $\left\{\hat{\mathcal{L}}_{i},\hat{\mathcal{L}}'_{i},\hat{\mathbb{L}}_{i}\right\}$ then satisfy the algebra:
\begin{equation} \label{rearrange-L.1*}
[\hat{\mathcal{L}}_{i},\hat{\mathcal{L}}_{j}] = i\hbar \, \epsilon_{ij}{}^{k} \hat{\mathcal{L}}_{k} \, , \quad [\hat{\mathcal{L}}'_{i},\hat{\mathcal{L}}'_{j}] = i\beta \, \epsilon_{ij}{}^{k} \hat{\mathcal{L}}'_{k} \, , 
\end{equation}
\begin{equation} \label{rearrange-L.2*}
[\hat{\mathcal{L}}_{i},\hat{\mathcal{L}}'_{j}] = [\hat{\mathcal{L}}'_{i},\hat{\mathcal{L}}_{j}] = 0 \, , 
\end{equation}
\begin{equation} \label{rearrange-L.3*}
[\hat{\mathcal{L}}_{i},\hat{\mathbb{L}}_{j}] - [\hat{\mathcal{L}}_{j},\hat{\mathbb{L}}_{i}] = i\hbar \, \epsilon_{ij}{}^{k} \hat{\mathbb{L}}_{k} \, , 
\end{equation}
\begin{equation} \label{rearrange-L.4*}
[\hat{\mathcal{L}}'_{i},\hat{\mathbb{L}}_{j}] - [\hat{\mathcal{L}}'_{j},\hat{\mathbb{L}}_{i}] = i\beta \, \epsilon_{ij}{}^{k} \hat{\mathbb{L}}_{k} \, , 
\end{equation}
\begin{equation} \label{rearrange-L.5*}
[\hat{\mathbb{L}}_{i},\hat{\mathbb{L}}_{j}] = i\beta \, \epsilon_{ij}{}^{k}\hat{\mathcal{L}}_{k} + i\hbar \, \epsilon_{ij}{}^{m}\hat{\mathcal{L}}'_{m} \, ,
\end{equation}
%
This is less restrictive than Equations (\ref{rearrange-L.1})-(\ref{rearrange-L.8}) since, together, Equations (\ref{rearrange-L.1})-(\ref{rearrange-L.8}) and (\ref{mathb{L}_{i}}) imply Equations (\ref{rearrange-L.1*})-(\ref{rearrange-L.5*}), but the converse statement does not hold. 

Thus, we may in principle construct an alternative set of operators $\left\{\hat{\mathcal{L}}_{i},\hat{\mathcal{L}}'_{i},\hat{\mathbb{L}}_{i}\right\}$, not defined by Equations (\ref{L_terms}) and (\ref{mathb{L}_{i}}), which nonetheless satisfy Equations (\ref{rearrange-L.1*})-(\ref{rearrange-L.5*}). 
In this paper, we do not investigate alternative solutions of either (\ref{rearrange-L.1})-(\ref{rearrange-L.8}) or (\ref{rearrange-L.1*})-(\ref{rearrange-L.5*}) in detail. 
However, we note that, keeping our previous definitions of $\hat{\mathcal{L}}_{i}$, $\hat{\mathcal{L}}'_{i}$ (\ref{L_terms}), and~defining the new operators $\hat{\mathbb{L}}_{i} := \frac{2}{\sqrt{\hbar\beta}}\epsilon_{i}{}^{jk}\hat{\mathcal{L}}_{j}\hat{\mathcal{L}}'_{k}$ (**), we may satisfy (\ref{rearrange-L.1*})--(\ref{rearrange-L.4*}) but not (\ref{rearrange-L.5*}).  

The operators (**) are not equivalent to those defined in Equation (\ref{mathb{L}_{i}}) and do not fully satisfy the generalised angular momentum algebra. 
Despite this, they offer an important clue about generalised spin physics in the smeared-space model, which will be considered in detail in the next section. 
As we will show, explicitly, in Section \ref{Sec.5.2}, it is straightforward to construct finite-dimensional analogues of the subcomponents $\left\{\hat{\mathcal{L}}_{i}\right\}_{i=1}^{3}$ and $\left\{\hat{\mathcal{L}}'_{i}\right\}_{i=1}^{3}$. 
However, it it is less obvious how to construct spin-operator counterparts of the commuting components $\left\{\hat{\Lambda}_i\right\}_{i=1}^{3}$ and $\left\{\hat{\Lambda}'_i\right\}_{i=1}^{3}$. 
Nonetheless, simple spin-operator analogues of $\left\{\hat{\mathbb{L}}_{i}\right\}_{i=1}^{3}$ exist. 
These take a form analogous to (**) with $\hat{\mathcal{L}}_{i}$ and $\hat{\mathcal{L}}'_{i}$ replaced by their finite-dimensional analogues. 
It may then be shown that, {\it if} the spin part of background state $\ket{g}$ is assumed to be fermionic, with eigenvalues $\pm \beta/2$, the resulting generalised spin operators satisfy all the relevant equations of a generalised spin algebra. 
This algebra has the same formal structure as Equations (\ref{rearrange-L.1*})-(\ref{rearrange-L.5*}).

Finally, we are now able to demonstrate that the generalised algebras (\ref{rearrange-L.1})-(\ref{rearrange-L.8}) and (\ref{rearrange-L.1*})-(\ref{rearrange-L.5*}) generate GURs for angular momentum. 
Depending on which algebra we choose, the uncertainties of the generalised angular momentum operators $\hat{L}_{i}$ (\ref{L_sum}) may be expressed in terms of the subcomponents $\left\{\hat{\mathcal{L}}_{i},\hat{\mathcal{L}}'_{i},\hat{\Lambda}_{i},\hat{\Lambda}'_{i}\right\}$ or $\left\{\hat{\mathcal{L}}_{i},\hat{\mathcal{L}}'_{i},\hat{\mathbb{L}}_{i}\right\}$, respectively. 

In terms of the first set of subcomponents, the variance of an individual component of the generalised angular momentum, $(\Delta_{\Psi}L_{i})^2$, is
\begin{eqnarray} \label{DL^2}
(\Delta_{\Psi}L_{i})^2 &=& (\Delta_{\Psi}\mathcal{L}_{i})^2 + (\Delta_{\Psi}\mathcal{L}'_{i})^2 + (\Delta_{\Psi}\Lambda_{i})^2 + (\Delta_{\Psi}\Lambda'_{i})^2
\nonumber\\
&+& {\rm cov}(\hat{\mathcal{L}}_{i},\hat{\Lambda}_{i}) + {\rm cov}(\hat{\Lambda}_{i},\hat{\mathcal{L}}_{i})
\nonumber\\
&+& {\rm cov}(\hat{\mathcal{L}}_{i},\hat{\Lambda}'_{i}) + {\rm cov}(\hat{\Lambda}'_{i},\hat{\mathcal{L}}_{i})
\nonumber\\
&+& {\rm cov}(\hat{\mathcal{L}}'_{i},\hat{\Lambda}_{i}) + {\rm cov}(\hat{\Lambda}_{i},\hat{\mathcal{L}}'_{i})
\nonumber\\
&+& {\rm cov}(\hat{\mathcal{L}}'_{i},\hat{\Lambda}'_{i}) + {\rm cov}(\hat{\Lambda}'_{i},\hat{\mathcal{L}}'_{i})
\nonumber\\
&+& {\rm cov}(\hat{\Lambda}_{i},\hat{\Lambda}'_{i}) + {\rm cov}(\hat{\Lambda}'_{i},\hat{\Lambda}_{i}) \, ,
\end{eqnarray}
since ${\rm cov}(\hat{\mathcal{L}}_{i},\hat{\mathcal{L}}'_{i}) = {\rm cov}(\hat{\mathcal{L}}'_{i},\hat{\mathcal{L}}_{i}) = 0$. 
The first term on the right-hand side represents the contribution to the total uncertainty from the canonical QM wave function $\psi$, the second represents the pure geometric part (that is, the contribution from $g$), and~the additional contributions are generated by operators that cannot be decomposed as either $\hat{\mathbb{1}} \otimes ( \dots )$ or $( \dots ) \otimes \hat{\mathbb{1}}$. 
Thus, Equation (\ref{DL^2}) takes a form analogous to Equations (\ref{comp-1}) and (\ref{comp-2}) but with additional cross terms, i.e., terms generated by operators that do not act on one subspace of the composite state $\ket{\Psi} = \ket{\psi} \otimes \ket{g}$ (\ref{Psi_unitary_equiv}) alone. 

We recall that Equations (\ref{comp-1}) and (\ref{comp-2}) are equivalent to Equations (\ref{X_uncertainty}) and (\ref{P_uncertainty}) and that these generate the GUP and the EUP, respectively, in the smeared-space model. 
In the case of GURs for position and linear momentum, we were able to use a simple theorem about the structure of convolutions to obtain Equations (\ref{X_uncertainty}) and (\ref{P_uncertainty}), even when momentum space representation of $\ket{\Psi}$ was expressed in terms of the entangled basis $\ket{\vec{p} \, \vec{p}{\, '}}$ (\ref{mod_dB}). 
However, in the case of angular momentum, it was necessary to first express $\ket{\Psi}$ in terms of a separable basis (\ref{Psi_unitary_equiv}) and to define the corresponding ``split'' operators (\ref{ops_split_Q}) and (\ref{ops_split_Pi}), before the generalised uncertainties $(\Delta_{\Psi}L_{i})^2$ could be decomposed into canonical and non-canonical parts. 
With this in mind, we note that only the first term on the right-hand side of Equation (\ref{DL^2}) is present in canonical QM, i.e., $\Delta_{\Psi}\mathcal{L}_{i} \equiv \Delta_{\psi}l_{i}$, where $\hat{l}_i :=\epsilon_{ij}{}^{k}\hat{x}^j\hat{p}_k$ is the canonical angular momentum operator. 
All additional terms are non-canonical and are a direct consequence of the smearing procedure (\ref{smear_map}).

Multiplying Equation (\ref{DL^2}) by a similar expression for $(\Delta_{\Psi}L_{j})^2$, we obtain the GUR for orbital angular momentum implied by the smeared-space model. 
Though it is beyond the scope of this paper to investigate the consequences of this relation in detail, we note that it is of the general form: 
\begin{eqnarray} \label{}
(\Delta_{\Psi}L_{i})^2(\Delta_{\Psi}L_{j})^2 \geq \dots \geq \left(\frac{\hbar + \beta}{2}\right)^2|(\epsilon_{ij}{}^{k})^2\braket{\hat{L}_{k}}_{\Psi}^2| \, .
\end{eqnarray}
The leading contribution to the terms in the middle is of the form $(\Delta_{\Psi}\mathcal{L}_{i})^2(\Delta_{\Psi}\mathcal{L}_{j})^2 \geq (\hbar/2)^2|(\epsilon_{ij}{}^{k})^2\braket{\hat{\mathcal{L}}_{k}}_{\Psi}^2|$, which is equivalent to the canonical QM uncertainty relation (\ref{ll_UR}). 
Again, we emphasise that all additional terms are non-canonical and arise as a direct consequence of the smearing map (\ref{smear_map}).

In terms of the second set of subcomponents, $(\Delta_{\Psi}L_{i})^2$ may also be written as:
\begin{eqnarray} \label{DL^2*}
(\Delta_{\Psi}L_{i})^2 &=& (\Delta_{\Psi}\mathcal{L}_{i})^2 + (\Delta_{\Psi}\mathcal{L}'_{i})^2 + (\Delta_{\Psi}\mathbb{L}_{i})^2 
\nonumber\\
&+& {\rm cov}(\hat{\mathcal{L}}_{i},\hat{\mathbb{L}}_{i}) + {\rm cov}(\hat{\mathbb{L}}_{i},\hat{\mathcal{L}}_{i})
\nonumber\\
&+& {\rm cov}(\hat{\mathcal{L}}'_{i},\hat{\mathbb{L}}_{i}) + {\rm cov}(\hat{\mathbb{L}}_{i},\hat{\mathcal{L}}'_{i}) \, .
\end{eqnarray}
Multiplying by the equivalent expression for $(\Delta_{\Psi}L_{j})^2$, we obtain an alternative (and simpler) form of the GUR for smeared-space angular momentum. 

\section{Spin in Smeared-Space QM} \label{Sec.5}

\subsection{Historical Analogy as a Guide to Generalisation} \label{Sec.5.1}

To construct a mathematical model of spin measurements in smeared-space, we proceed by analogy with the historical development of canonical QM. 
(See Appendix \ref{Appendix.C.1} for details.) 
Hence, we seek a set of constant-valued matrices $\left\{\hat{S}_{i}\right\}_{i=1}^{3}$ that satisfy the same algebraic structures as the components of angular momentum $\left\{\hat{L}_{i}\right\}_{i=1}^{3}$. 

In the canonical theory, the relevant algebra for the angular momentum operators is simply the three-dimensional rotation algebra, ${\rm so}(3)$, scaled by a factor of $\hbar$ (\ref{ll_commutator}). 
However, in the smeared-space model, the situation is more complicated. 
In Section \ref{Sec.4.3}, we showed how the smeared-space angular momentum operators can be decomposed into the sum of four terms: a canonical quantum term $\hat{\mathcal{L}}_{i}$ acting on the first subspace of the tensor product state $\ket{\Psi}$ (\ref{Psi_unitary_equiv}), a ``pure'' geometric part $\hat{\mathcal{L}}'_{i}$ acting on the second, and~two ``cross terms'', $\hat{\Lambda}_{i}$ and $\hat{\Lambda}'_{i}$, acting on both subspaces (\ref{L_sum})--(\ref{L_terms}). 
The subcomponents $\left\{\hat{\mathcal{L}}_{i},\hat{\mathcal{L}}'_{i},\hat{\Lambda}_{i},\hat{\Lambda}'_{i}\right\}$ were found to obey the subalgebra defined by Equations (\ref{rearrange-L.1})--(\ref{rearrange-L.8}). 
Together, these equations ensure that the rescaled ${\rm so}(3)$ Lie algebra, with $\hbar \rightarrow \hbar + \beta$ (\ref{LL_commutator}), holds for $\left\{\hat{L}_{i}\right\}_{i=1}^{3}$. 
In~addition, we used the alternative definition $\hat{\mathbb{L}}_i := \hat{\Lambda}_{i} +  \hat{\Lambda}'_{i}$ (\ref{mathb{L}_{i}}), leading to the subalgebra (\ref{rearrange-L.1*})--(\ref{rearrange-L.5*}) for $\left\{\hat{\mathcal{L}}_{i},\hat{\mathcal{L}}'_{i},\hat{\mathbb{L}}_i\right\}$. 

Hence, when searching for generalised spin operators, whose eigenvalues are to be interpreted as the possible spins of the composite matter-plus-geometry quantum state, we have three possible options to explore. 
First, we may search for exact analogue Equations (\ref{L_sum}) and (\ref{L_terms}). 
This requires $\hat{S}_{i}$ to be decomposed into the sum of four terms, $\hat{S}_{i} = \hat{\mathcal{S}}_{i} + \hat{\mathcal{S}}'_{i} + \hat{\Sigma}_{i} + \hat{\Sigma}'_{i}$, where $\hat{\mathcal{S}}_{i} := \epsilon_{ij}{}^{k}\hat{\alpha}^{j}\hat{\beta}_{k}$, $\hat{\mathcal{S}}'_{i} := \epsilon_{ij}{}^{k}\hat{\alpha}'^{j}\hat{\beta}'_{k}$, $\hat{\Sigma}_{i} := \epsilon_{ij}{}^{k}\hat{\alpha}^{j}\hat{\beta}'_{k}$ and $\hat{\Sigma}'_{i} := \epsilon_{ij}{}^{k}\hat{\alpha}'^{j}\hat{\beta}_{k}$. 
In this case, $\hat{\alpha}^{i}$ and $\hat{\beta}_{j}$ are required to be finite-dimensional constant-valued matrices, acting on the first spin-subspace of the smeared tensor product state, that satisfy the $\hbar$-scaled Heisenberg algebra: $[\hat{\alpha}^{i},\hat{\beta}_{j}] = i\hbar\delta^{i}{}_{j} \, \hat{\mathbb{I}}$, $[\hat{\alpha}^{i},\hat{\alpha}^{j}] = 0$, $[\hat{\beta}_{i},\hat{\beta}_{j}] = 0$. 
(Here, $\hat{\mathbb{I}}$ is used to denote the tensor product of the two spin subspaces, corresponding to matter and geometry, respectively.)
Similarly, $\hat{\alpha}'^{i}$ and $\hat{\beta}'_{j}$ must be finite-dimensional constant-valued matrices, acting on the second subspace of the tensor product, satisfying the $\beta$-scaled Heisenberg algebra: $[\hat{\alpha}'^{i},\hat{\beta}'_{j}] = i\beta\delta^{i}{}_{j} \, \hat{\mathbb{I}}$, $[\hat{\alpha}'^{i},\hat{\alpha}'^{j}] = 0$, $[\hat{\beta}'_{i},\hat{\beta}'_{j}] = 0$. 
(The requirement that each representation of the Heisenberg algebra acts on a different subspace of the product state also ensures that $[\hat{\alpha}^{i},\hat{\alpha}'^{j}] = 0$, $[\hat{\beta}_{i},\hat{\beta}'_{j}] = 0$, $[\hat{\alpha}_{i},\hat{\beta}'_{j}] = 0$ and $[\hat{\alpha}'_{i},\hat{\beta}_{j}] = 0$.)

However, it is straightforward to show that no such matrices exist. 
The matrices most similar to those we require are finite-dimensional representations of the Heisenberg group \cite{HeisenbergGroup}. 
This group has one central element ($z$) and two sets of generators, usually denoted $x^{i}$ and $p_{j}$ by analogy with the canonical commutation relations, that satisfy the following algebra: $[x^{i},p_{j}] = \delta^{i}{}_{j} \, z$, $[x^{i},x^{j}] = 0$, $[p_{i},p_{j}] = 0$ and $[x^{i},z] = [z,x^{i}]$, $[p_{j},z] = [z,p_{j}]$. 
In other words, while the central element $z$ commutes with all other generators, it is {\it not} the identity element. 
Perhaps confusingly, the previous commutation relations are also typically referred to as the ``Heisenberg algebra'' in the mathematical literature, since they are the algebra of the Heisenberg group. 
However, they are {\it not} equivalent to the position-momentum commutation relations of canonical QM \cite{HeisenbergGroup}. 
Therefore, this procedure fails, as it is impossible to define exact finite-dimensional analogues of the $\hat{L}_i$ subcomponents $\left\{\hat{\mathcal{L}}_{i},\hat{\mathcal{L}}'_{i},\hat{\Lambda}_{i},\hat{\Lambda}'_{i}\right\}$.

Second, we may search for an alternative set of finite-dimensional constant-valued matrices, $\left\{\hat{\mathcal{S}}_{i},\hat{\mathcal{S}}'_{i},\hat{\Sigma}_{i},\hat{\Sigma}'_{i}\right\}$, satisfying the relevant algebra. 
By the argument above, these cannot be defined in terms of finite-dimensional analogues of the position and momentum operators, i.e., $\hat{\alpha}^{i} \sim \hat{X}^{i}$, $\hat{\alpha}'^{i} \sim \hat{X}'^{i}$, etc.
Formally, we require $\left\{\hat{\mathcal{S}}_{i},\hat{\mathcal{S}}'_{i},\hat{\Sigma}_{i},\hat{\Sigma}'_{i}\right\}$ to satisfy an algebra analogous to (\ref{rearrange-L.1})-(\ref{rearrange-L.8}) under the interchange $\hat{\mathcal{S}}_{i} \leftrightarrow \hat{\mathcal{L}}_{i}$, $\hat{\mathcal{S}}'_{i} \leftrightarrow \hat{\mathcal{L}}'_{i}$, $\hat{\Sigma}_{i} \leftrightarrow \hat{\Lambda}_{i}$ and $\hat{\Sigma}'_{i} \leftrightarrow \hat{\Lambda}'_{i}$. 
In this case, we must again require that $\hat{\mathcal{S}}_{i}$ act on the first subspace of the tensor product state, that $\hat{\mathcal{S}}'_{i}$ act on the second subspace, and~that $\hat{\Sigma}_{i}$ and $\hat{\Sigma}'_{i}$ act on both subspaces simultaneously. 
With this in mind, we note that the most natural operator that is able to act nontrivially on both spin subspaces is of the form $\hat{\Sigma}_{i} \sim \hat{\Sigma}'_{i} \sim \epsilon_{i}{}^{jk}\sigma_{j} \otimes \sigma'_{k}$. 
However, it is straightforward to show that, using this definition, $[\hat{\Sigma}_{i}, \hat{\Sigma}_{j}] \neq 0$ and $[\hat{\Sigma}'_{i}, \hat{\Sigma}'_{j}] \neq 0$, so that the analogues of Equations (\ref{rearrange-L.7}) cannot be satisfied. 
Therefore, this procedure also fails. 

Third, we may search for a smaller set of finite-dimensional constant-valued matrices, $\left\{\hat{\mathcal{S}}_{i},\hat{\mathcal{S}}'_{i},\hat{\mathbb{S}}_{i}\right\}$, satisfying an analogue of the algebra (\ref{rearrange-L.1*})-(\ref{rearrange-L.5*}) under the exchange $\hat{\mathcal{S}}_{i} \leftrightarrow \hat{\mathcal{L}}_{i}$, $\hat{\mathcal{S}}'_{i} \leftrightarrow \hat{\mathcal{L}}'_{i}$ and $\hat{\mathbb{S}}_{i} \leftrightarrow \hat{\mathbb{L}}_{i}$. 
Based on our previous considerations, this is clearly the most promising route. 
In the following section, we explore this possibility and construct explicit representations of the generator subcomponents $\hat{\mathcal{S}}_{i}$, $\hat{\mathcal{S}}'_{i}$ and $\hat{\mathbb{S}}_{i}$. 

\subsection{Generalised Algebra and GURs} \label{Sec.5.2}

Considering the arguments presented above, we define the generalised spin operator $\hat{S}_i$ as
\begin{eqnarray} \label{S_sum}
\hat{S}_{i} = \hat{\mathcal{S}}_{i} + \hat{\mathcal{S}}'_{i} + \hat{\mathbb{S}}_{i} \, , 
\end{eqnarray}
where $\hat{\mathcal{S}}_{i}$ and $\hat{\mathcal{S}}'_{i}$ are given by
\begin{eqnarray} \label{ss'}
\hat{\mathcal{S}}_{i} := \hat{s}_{i} \otimes \hat{\mathbb{1}}' \, , \quad \hat{\mathcal{S}}'_{i} := \hat{\mathbb{1}} \otimes \hat{s}'_{i} \, , 
\end{eqnarray} 
and
\begin{eqnarray} \label{s'}
\hat{s}_{i} := \frac{\hbar}{2} \, \sigma_{i} \, , \quad \hat{s}'_{i} := \frac{\beta}{2} \, \sigma'_{i} \, .
\end{eqnarray}
Equations (\ref{s'}) represent an extension of Equation (\ref{s}), which holds only for the matter sector.
The prime on the Pauli operators acting on the second spin-subspace, corresponding to the spin part of the quantum state associated with the background geometry, indicates that this may posses a different fundamental spin 
to the matter component, $s' \neq s$. 
In this case, the two spin subspaces have different dimensions. 
(In other words, we use the shorthand notations $\sigma'_{i} = \sigma_{i}(s')$, $\hat{\mathbb{1}}' = \hat{\mathbb{1}}_{2s'+1}$ and $\sigma_{i} = \sigma_{i}(s)$, $\hat{\mathbb{1}} = \hat{\mathbb{1}}_{2s+1}$.)
It follows from the definitions (\ref{S_sum})--(\ref{s'}) that
\begin{equation} \label{rearrange-S.1*}
[\hat{\mathcal{S}}_{i},\hat{\mathcal{S}}_{j}] = i\hbar \, \epsilon_{ij}{}^{k} \hat{\mathcal{S}}_{k} \, ,  \quad [\hat{\mathcal{S}}'_{i},\hat{\mathcal{S}}'_{j}] = i\beta \, \epsilon_{ij}{}^{k} \hat{\mathcal{S}}'_{k} \, , 
\end{equation}
and
\begin{equation} \label{rearrange-S.2*}
[\hat{\mathcal{S}}_{i},\hat{\mathcal{S}}'_{j}] = [\hat{\mathcal{S}}'_{i},\hat{\mathcal{S}}_{j}] = 0 \, ,
\end{equation}
for any $s$, $s'$.

Next, we define $\hat{\mathbb{S}}_{i}$ as
\begin{eqnarray} \label{mathbb{S}_{i}}
\hat{\mathbb{S}}_{i} &:=& \frac{\sqrt{\hbar\beta}}{2} \, \epsilon_{i}{}^{jk}\sigma_{j} \otimes \sigma'_{k} 
\nonumber\\
&=& \frac{2}{\sqrt{\hbar\beta}}  \, \epsilon_{i}{}^{jk} \hat{\mathcal{S}}_{j}\hat{\mathcal{S}}'_{k} \, . 
\end{eqnarray}
This is clearly the analogue of the operator (**) introduced below Equations (\ref{rearrange-L.1*})-(\ref{rearrange-L.5*}). 
Using the identity $[AB,C] = A[B,C] + [A,C]B$, Equations (\ref{rearrange-S.1*})--(\ref{mathbb{S}_{i}}) are sufficient to show that the relations
\begin{equation} \label{rearrange-S.3*}
[\hat{\mathcal{S}}_{i},\hat{\mathbb{S}}_{j}] - [\hat{\mathcal{S}}_{j},\hat{\mathbb{S}}_{i}] = i\hbar \, \epsilon_{ij}{}^{k} \hat{\mathbb{S}}_{k} \, , 
\end{equation}
\begin{equation} \label{rearrange-S.4*}
[\hat{\mathcal{S}}'_{i},\hat{\mathbb{S}}_{j}] - [\hat{\mathcal{S}}'_{j},\hat{\mathbb{S}}_{i}] = i\beta \, \epsilon_{ij}{}^{k} \hat{\mathbb{S}}_{k} \, , 
\end{equation}
also hold for any values of $s$ and $s'$. 
Hence, in order to recover a rescaled spin Lie algebra for the generalised operators $\left\{\hat{S}_i\right\}_{i=1}^{3}$ (with $\hbar \rightarrow \hbar + \beta$), we require the following commutation relations to hold between the cross terms 
$\hat{\mathbb{S}}_{i}$ and $\hat{\mathbb{S}}_{j}$: 
\begin{equation} \label{rearrange-S.5*}
[\hat{\mathbb{S}}_{i},\hat{\mathbb{S}}_{j}] = i\beta \, \epsilon_{ij}{}^{k}\hat{\mathcal{S}}_{k} + i\hbar \, \epsilon_{ij}{}^{m}\hat{\mathcal{S}}'_{m} \, .
\end{equation}

In this section, our main aim is to describe the generalised spin physics of electrons in smeared-space. 
Hence, since the situation in which $s = 1/2$ is of greatest physical interest, we restrict ourselves to this from now on. 
We then have
\begin{eqnarray} \label{fundamental_relation_spin_ops*}
\hat{\mathcal{S}}_{i}\hat{\mathcal{S}}_{j} = \left(\frac{\hbar}{2}\right)^2 \, \delta_{ij} \hat{\mathbb{I}}  + \, i\left(\frac{\hbar}{2}\right)\epsilon_{ij}{}^{k}\hat{\mathcal{S}}_{k} \, , 
\end{eqnarray}
and
\begin{eqnarray} \label{ss_anti-commutator*}
[\hat{\mathcal{S}}_{i},\hat{\mathcal{S}}_{j}]_{+} = \frac{\hbar^2}{2} \, \delta_{ij} \, \hat{\mathbb{I}} \, ,
\end{eqnarray}
where $[ \, . \, , \, . \, ]_{+}$ denotes the anti-commutator, which are equivalent to the canonical Equations (\ref{fundamental_relation_spin_ops}) and (\ref{ss_anti-commutator}), respectively. 
It is then straightforward to show that Equation (\ref{rearrange-S.5*}) holds {\it if and only if}
\begin{eqnarray} \label{fundamental_relation_spin_ops**}
\hat{\mathcal{S}}'_{i}\hat{\mathcal{S}}'_{j} = \left(\frac{\beta}{2}\right)^2 \, \delta_{ij} \hat{\mathbb{I}}  + \, i\left(\frac{\beta}{2}\right)\epsilon_{ij}{}^{k}\hat{\mathcal{S}}'_{k} \, , 
\end{eqnarray}
so that
\begin{eqnarray} \label{ss_anti-commutator**}
[\hat{\mathcal{S}}'_{i},\hat{\mathcal{S}}'_{j}]_{+} = \frac{\beta^2}{2} \, \delta_{ij} \, \hat{\mathbb{I}} \, .
\end{eqnarray}
However, unlike Equations (\ref{rearrange-S.1*})--(\ref{rearrange-S.2*}) and (\ref{rearrange-S.3*})--(\ref{rearrange-S.4*}), these relations hold only for $s' = 1/2$. 
(See Appendix \ref{Appendix.C.2}.)
Hence, consistency of the generalised spin structure implies that the quantum state associated with the background geometry must be {\it fermionic}, with spin values $\pm\beta/2$.  

The generalised spin algebra for the subcomponents $\left\{\hat{\mathcal{S}}_{i},\hat{\mathcal{S}}'_{i},\hat{\mathbb{S}}_{i}\right\}$ is, therefore
\begin{equation} \label{rearrange-S.1}
[\hat{\mathcal{S}}_{i},\hat{\mathcal{S}}_{j}] = i\hbar \, \epsilon_{ij}{}^{k} \hat{\mathcal{S}}_{k} \, ,  \quad [\hat{\mathcal{S}}'_{i},\hat{\mathcal{S}}'_{j}] = i\beta \, \epsilon_{ij}{}^{k} \hat{\mathcal{S}}'_{k} \, , 
\end{equation}
\begin{equation} \label{rearrange-S.2}
[\hat{\mathcal{S}}_{i},\hat{\mathcal{S}}'_{j}] = [\hat{\mathcal{S}}'_{i},\hat{\mathcal{S}}_{j}] = 0 \, , 
\end{equation}
\begin{equation} \label{rearrange-S.3}
[\hat{\mathcal{S}}_{i},\hat{\mathbb{S}}_{j}] - [\hat{\mathcal{S}}_{j},\hat{\mathbb{S}}_{i}] = i\hbar \, \epsilon_{ij}{}^{k} \hat{\mathbb{S}}_{k} \, , 
\end{equation}
\begin{equation} \label{rearrange-S.4}
[\hat{\mathcal{S}}'_{i},\hat{\mathbb{S}}_{j}] - [\hat{\mathcal{S}}'_{j},\hat{\mathbb{S}}_{i}] = i\beta \, \epsilon_{ij}{}^{k} \hat{\mathbb{S}}_{k} \, , 
\end{equation}
\begin{equation} \label{rearrange-S.5}
[\hat{\mathbb{S}}_{i},\hat{\mathbb{S}}_{j}] = i\beta \, \epsilon_{ij}{}^{k}\hat{\mathcal{S}}_{k} + i\hbar \, \epsilon_{ij}{}^{m}\hat{\mathcal{S}}'_{m} \, ,
\end{equation}
%
Together, Equations (\ref{rearrange-S.1})--(\ref{rearrange-S.5}) give rise to the rescaled ${\rm su}(2)$ Lie algebra:
\begin{eqnarray} \label{SS_commutator}
[\hat{S}_{i},\hat{S}_{j}] = i(\hbar + \beta) \epsilon_{ij}{}^{k}\hat{S}_{k} \, ,
\end{eqnarray}
and the rescaled Clifford algebra:
\begin{eqnarray} \label{SS_anticommutator}
[\hat{S}_{i},\hat{S}_{j}]_{+} = \frac{(\hbar + \beta)^2}{2} \delta_{ij} \, \hat{\mathbb{I}} \, ,
\end{eqnarray}
for the generalised spin-measurement operators $\left\{\hat{S}_{i}\right\}_{i=1}^{3}$ (\ref{S_sum}). 
From (\ref{SS_commutator}), it also follows that
\begin{eqnarray} \label{}
[\hat{S}^{2},\hat{S}_{i}] = 0 \, . 
\end{eqnarray}

Note that, in the limit $\hbar \rightarrow \beta$, the $\hat{\mathbb{S}}_{i}$ term is not necessary to maintain the canonical Lie algebra structure. 
Since both $\left\{\hat{\mathcal{S}}_{i}\right\}_{i=1}^{3}$ and $\left\{\hat{\mathcal{S}}'_{i}\right\}_{i=1}^{3}$ are representations of the ${\rm su}(2)$ generators, and~these representations commute with each other (\ref{ss'}), the combination 
$\hat{\mathcal{S}}_{i} + \hat{\mathcal{S}}'_{i} =: \hat{S}_{i}$ also satisfies the ${\rm su}(2)$ algebra {\it if} both sets of generators are weighted by the same scale factor. 
In this case, we may pull a single factor of $\hbar$ outside the sum of terms on right-hand sides of the commutation relations, yielding $[\hat{S}_{i},\hat{S}_{j}] = i \hbar\epsilon_{ij}{}^{k}(\hat{\mathcal{S}}_{k} + \hat{\mathcal{S}}'_{k}) =: i\hbar \epsilon_{ij}{}^{k}\hat{S}_{i}$. 
However, in the presence of a two-scale theory, which is an essential feature of the smeared-space model \cite{Lake:2018zeg}, the presence of $\hat{\mathbb{S}}_{i}$ is unavoidable. 
Without it, it is not possible to construct an operator $\hat{S}_{i}$ that includes commuting representations of ${\rm su}(2)$ weighted by different scale factors, i.e., $\hat{\mathcal{S}}_{i} = (\hbar/2) (\sigma_{i} \otimes \mathbb{1}')$ and $\hat{\mathcal{S}}'_{i} = (\beta/2) (\mathbb{1} \otimes \sigma'_{i})$ ($\beta \neq \hbar$), and~which also satisfies a canonical-type commutation relation. 
In this case, it is not possible to pull a single factor (with units of action) outside the expression on the right-hand side of the relation $[\hat{S}_{i},\hat{S}_{j}] = (\dots)$ without including $\hat{\mathbb{S}}_{i}$ (\ref{mathbb{S}_{i}}) in the definition of $\hat{S}_{i}$ (\ref{S_sum}). 

This is a fundamental difference between canonical two-particle states and the bipartite matter-plus-geometry states of the smeared-space model. 
Furthermore, it has clear physical interpretation. 
The first copy of the ${\rm su}(2)$ algebra, weighted by $\hbar$, is generated by the spin of the matter sector, whereas the second copy, weighted by $\beta$, is generated by the intrinsic spin of the background. 
If these spins are left to evolve freely, {\it without interacting}, the introduction of a second quantisation scale for geometry, $\beta \neq \hbar$, breaks the ${\rm SU}(2)$ invariance of the {\it composite} matter-plus-geometry state. 
However, the spins do not evolve freely but interact via the cross term $\hat{\mathbb{S}}_{i}$. 
The interaction is such that ${\rm SU}(2)$ symmetry is restored, for the composite state, under a simple rescaling $\hbar \rightarrow \hbar + \beta$.

Written explicitly, the generalised spin matrices take the form:
\begin{eqnarray} \label{S_i_explicit}
&&\hat{S}_{x} = 
\begin{bmatrix}
    0  &  \frac{(\beta + i\sqrt{\hbar\beta})}{2} & \frac{(\hbar - i\sqrt{\hbar\beta})}{2}  &  0 \\
    \frac{(\beta - i\sqrt{\hbar\beta})}{2}  &  0 & 0  &  \frac{(\hbar + i\sqrt{\hbar\beta})}{2} \\
    \frac{(\hbar + i\sqrt{\hbar\beta})}{2} &  0 & 0  &  \frac{(\beta - i\sqrt{\hbar\beta})}{2} \\
    0  &  \frac{(\hbar - i\sqrt{\hbar\beta})}{2}  & \frac{(\beta + i\sqrt{\hbar\beta})}{2} &  0 
\end{bmatrix}
\, , \nonumber\\ 
&&\hat{S}_{y} = 
\begin{bmatrix}
    0  &  -\frac{(i\beta - \sqrt{\hbar\beta})}{2} & -\frac{(i\hbar + \sqrt{\hbar\beta})}{2}  &  0 \\
    \frac{(i\beta + \sqrt{\hbar\beta})}{2}  &  0 & 0  &  -\frac{(i\hbar - \sqrt{\hbar\beta})}{2} \\
    \frac{(i\hbar - \sqrt{\hbar\beta})}{2} &  0 & 0  &  -\frac{(i\beta + \sqrt{\hbar\beta})}{2} \\
    0  &  \frac{(i\hbar + \sqrt{\hbar\beta})}{2}  & \frac{(i\beta - \sqrt{\hbar\beta})}{2}  &  0 
\end{bmatrix}
\, , \nonumber\\
&&\hat{S}_{z} =
\begin{bmatrix}
    \frac{(\hbar + \beta)}{2}  &  0  &  0  &  0 \\
    0  &  \frac{(\hbar - \beta)}{2}  &  i\sqrt{\hbar\beta}  &  0 \\
    0  &  -i\sqrt{\hbar\beta}  &  -\frac{(\hbar - \beta)}{2}  &  0 \\
    0  &  0  &  0  &  -\frac{(\hbar+\beta)}{2} 
\end{bmatrix}
\, ,
\end{eqnarray}
and $\hat{S}^2$ is given by
\begin{eqnarray}  \label{S^2_explicit}
\hat{S}^{2} = \frac{3(\hbar + \beta)^2}{4} \, \hat{\mathbb{1}}_4 \, . 
\end{eqnarray}
This follows from the fact that the matrices $\left\{\left(\frac{\hbar + \beta}{2}\right)^{-1}\hat{S}_{i}\right\}_{i=1}^{3}$ are involutions. 
Hence, in the smeared-space model, $\left\{\left(\frac{\hbar + \beta}{2}\right)^{-1}\hat{S}_{i}\right\}_{i=1}^{3}$ are the analogues of the canonical spin-$1/2$ Pauli matrices, $\left\{\sigma_{i}\right\}_{i=1}^{3} = \left\{\left(\frac{\hbar}{2}\right)^{-1}\hat{s}_{i}\right\}_{i=1}^{3}$. 
However, unlike the canonical Pauli matrices, $\left\{\left(\frac{\hbar + \beta}{2}\right)^{-1}\hat{S}_{i}\right\}_{i=1}^{3}$ depend explicitly on both quantisation scales, $\hbar$ and $\beta$.

It is straightforward to verify that all three spin operators $\left\{\hat{S}_i\right\}_{i=1}^{3}$ (\ref{S_i_explicit}) have the eigenvalues:
\begin{eqnarray}
\left\{ \frac{(\hbar + \beta)}{2} ,  \frac{(\hbar + \beta)}{2} ,  -\frac{(\hbar + \beta)}{2} ,  -\frac{(\hbar + \beta)}{2} \right\} \, , 
\end{eqnarray}
which, for $\hat{S}_z$, correspond to the following (un-normalised) eigenvectors:
\begin{eqnarray}
\left\{(1,0,0,0),\left(0,\frac{i\hbar}{\sqrt{\hbar\beta}},1,0\right),\left(0,-\frac{i\beta}{\sqrt{\hbar\beta}},1,0\right), (0,0,0,1)\right\}.
\nonumber
\end{eqnarray}

The normalised eigenvectors of $\hat{S}_{z}$ may then be written as
\begin{eqnarray} \label{unentangled_eigenvectors}
\Big|\frac{3(\hbar + \beta)^2}{4}, \frac{(\hbar + \beta)}{2}\Big\rangle &=& (1,0,0,0) = \ket{\uparrow}_1 \ket{\uparrow}_2 \, ,  
\nonumber\\ 
\Big|\frac{3(\hbar + \beta)^2}{4}, -\frac{(\hbar + \beta)}{2}\Big\rangle &=& (0,0,0,1) = \ket{\downarrow}_1 \ket{\downarrow}_2  \, ,  
\nonumber\\
\end{eqnarray}
and
\begin{eqnarray} \label{entangled_eigenvectors}
&&\Big|\frac{3(\hbar + \beta)^2}{4}, \frac{(\hbar + \beta)}{2}\Big\rangle_{\delta} = \frac{1}{\sqrt{1 + \delta}} (0,1,-i\sqrt{\delta},0)
\nonumber\\
&=& \frac{1}{\sqrt{1 + \delta}}(\ket{\uparrow}_1 \ket{\downarrow}_2 - i\sqrt{\delta}\ket{\downarrow}_1 \ket{\uparrow}_2) \, , 
\nonumber\\
&&\Big|\frac{3(\hbar + \beta)^2}{4}, -\frac{(\hbar + \beta)}{2}\Big\rangle_{\delta} = \frac{1}{\sqrt{1 + \delta}} (0,i\sqrt{\delta},1,0) 
\nonumber\\ 
&=& \frac{1}{\sqrt{1 + \delta}} (\ket{\downarrow}_1 \ket{\uparrow}_2 + i\sqrt{\delta}\ket{\uparrow}_1 \ket{\downarrow}_2)  \, ,
\end{eqnarray}
where 
\begin{eqnarray}
\delta := \hbar/\beta \simeq 10^{-61} \, . 
\end{eqnarray}

Hence, the single electron plus smeared-background system has {\it four} spin states, as opposed to the two spin states of electrons on the fixed background of canonical QM. 
However, the operators $\hat{S}^{2}$ and $\hat{S}_{z}$ that act on the composite system have only two distinct sets of eigenvalues, $\left\{3(\hbar+\beta)^2/4,\pm(\hbar+\beta)/2\right\}$. 
Each pair of eigenvalues has a $2$-fold degeneracy, corresponding to one separable state and one state in which the spins of the electron and the background are entangled.  
The eigenvectors $\ket{\frac{3(\hbar + \beta)^2}{4}, \frac{(\hbar + \beta)}{2}}$ and $\ket{\frac{3(\hbar + \beta)^2}{4}, \frac{(\hbar + \beta)}{2}}_{\delta}$ correspond to spin ``up'' states, according to the measured values of $\hat{S}^{2}$ and $\hat{S}_{z}$, whereas $\ket{\frac{3(\hbar + \beta)^2}{4}, -\frac{(\hbar + \beta)}{2}}$ and $\ket{\frac{3(\hbar + \beta)^2}{4}, -\frac{(\hbar + \beta)}{2}}_{\delta}$ correspond to spin ``down'' states. 

For the unentangled states, $\ket{\frac{3(\hbar + \beta)^2}{4}, \pm \frac{(\hbar + \beta)}{2}}$, the spins of the matter and geometry components of the tensor-product smeared-state, $\ket{\psi}$ and $\ket{g}$, are aligned. 
The spin up state is characterised by the individual values $\left\{\hbar/2,\beta/2\right\}$ and the spin down state by the values $\left\{-\hbar/2,-\beta/2\right\}$. 
However, for the entangled eigenvectors, $\ket{\frac{3(\hbar + \beta)^2}{4}, \pm \frac{(\hbar + \beta)}{2}}_{\delta}$, there is no simple relation between the matter and geometry components of the total quantum state. 
Remarkably, the entangled eigenstates (\ref{entangled_eigenvectors}) have the same eigenvalues as the simple separable states (\ref{unentangled_eigenvectors}). 

We also note that, in the absence the interaction term $\mathbb{S}_{i}$, the eigenvalues of the composite operator $\hat{\mathcal{S}}_{i} + \hat{\mathcal{S}}'_{i}$ are $\left\{(\hbar+\beta)/2,(\hbar-\beta)/2,(-\hbar+\beta)/2,-(\hbar+\beta)/2\right\}$. 
These correspond to the eigenvectors $\left\{\ket{\uparrow}_1 \ket{\uparrow}_2,\ket{\uparrow}_1 \ket{\downarrow}_2,\ket{\downarrow}_1 \ket{\uparrow}_2,\ket{\downarrow}_1 \ket{\downarrow}_2\right\}$, respectively, which in the limit $\beta \rightarrow \hbar$ yield the familiar spin eigenvectors of a canonical two-particle state \cite{Rae}.  
Thus, the introduction of $\mathbb{S}_{i}$ not only restores ${\rm SU}(2)$ symmetry in the composite matter-plus-geometry system, in the presence of a two-scale theory with $\beta \neq \hbar$, but also alters {\it two} of the four spin-eigenstates while leaving the remaining two unchanged. 
This, in turn, shifts the corresponding eigenvalues by {\it just} the right amount to introduce $2$-fold degeneracy in the measured values of $\hat{S}^{2}$ and $\hat{S}_{z}$. 

A priori, there was no reason for us to anticipate that the additional terms required to restore ${\rm SU}(2)$ symmetry, i.e., those involving $\mathbb{S}_{i}$ in the algebra (\ref{rearrange-S.1})-(\ref{rearrange-S.5}), would simultaneously introduce degeneracy of the resulting spin states.  
However, had this {\it not} been the case, the doubling of the spin degrees of freedom would, in principle, have been directly detectable via simultaneous measurements of $\hat{S}^2$ and $\hat{S}_z$. 
This would have caused severe problems for the smeared-space model, at least philosophically, even if no mathematical inconsistencies were introduced. 
It is straightforward to see why.

In the non-spin part of the model, the doubling of the canonical degrees of freedom is detectable only indirectly, via the additional statistical fluctuations it induces in the measured values of position, momentum and angular momentum, etc. 
These generate the GURs discussed in previous sections. 
In effect, we assume a measurement scheme in which measurements can be made only on material bodies in space \cite{Lake:2018zeg}. 
Hence, we do not have direct physical access to the quantum degrees of freedom of the background, which can be detected only indirectly via their influence on quantum particles. 
Mathematically, this is expressed by tracing out, or, equivalently, integrating out the degrees of freedom in the first subspace of the tensor product Hilbert space, as in Equation (\ref{EQ_XPRIMEDENSITY}). 
However, since the spin part of the composite matter-plus-geometry state is finite-dimensional, no integrals appear anywhere in the corresponding formulae. 
Furthermore, there is no clear physical justification for tracing out half of the doubled spin degrees of freedom, since this would require us to make an arbitrary choice, i.e., which {\it two} of the four possible spin states should we regard as physical?

Remarkably, the algebra (\ref{rearrange-S.1})-(\ref{rearrange-S.5}) saves us from this dilemma, just as it ``saves'' the ${\rm SU}(2)$ symmetry of the two-scale quantisation scheme. 
The resulting generalised spin model is both mathematically consistent {\it and} consistent with the physical assumptions underlying the smeared-space model as a whole, despite the doubling of the number of dimensions in the spin Hilbert space.

Finally, we consider the GURs implied by the generalised spin algebra (\ref{rearrange-S.1})-(\ref{rearrange-S.5}). 
By analogy with Equation (\ref{DL^2*}), $(\Delta_{\Psi}S_{i})^2$ takes the form: 
\begin{eqnarray} \label{DS^2*}
(\Delta_{\Psi}S_{i})^2 &=& (\Delta_{\Psi}\mathcal{S}_{i})^2 + (\Delta_{\Psi}\mathcal{S}'_{i})^2 + (\Delta_{\Psi}\mathbb{S}_{i})^2 
\nonumber\\
&+& {\rm cov}(\hat{\mathcal{S}}_{i},\hat{\mathbb{S}}_{i}) + {\rm cov}(\hat{\mathbb{S}}_{i},\hat{\mathcal{S}}_{i})
\nonumber\\
&+& {\rm cov}(\hat{\mathcal{S}}'_{i},\hat{\mathbb{S}}_{i}) + {\rm cov}(\hat{\mathbb{S}}_{i},\hat{\mathcal{S}}'_{i}) \, .
\end{eqnarray}
Multiplying by the equivalent expression for $(\Delta_{\Psi}S_{j})^2$, we obtain the GUR for spin measurements in smeared-space. 
Again, it is beyond the scope of this paper to investigate the consequences of this relation in detail. 
Nonetheless, we note that it is of the general form: 
\begin{eqnarray} 
(\Delta_{\Psi}S_{i})^2(\Delta_{\Psi}S_{j})^2 \geq \dots \geq \left(\frac{\hbar + \beta}{2}\right)^2|(\epsilon_{ij}{}^{k})^2\braket{\hat{S}_{k}}_{\Psi}^2| \, ,
\end{eqnarray}
where the leading contribution to the terms in the middle is of the form $(\Delta_{\Psi}\mathcal{S}_{i})^2(\Delta_{\Psi}\mathcal{S}_{j})^2 \geq (\hbar/2)^2|(\epsilon_{ij}{}^{k})^2\braket{\hat{\mathcal{S}}_{k}}_{\Psi}^2|$. 
This is equivalent to the canonical uncertainty relation for spin measurements. 
The additional terms are non-canonical and depend on the ratio of the dark energy density to the Planck density, which determines the value of the geometry quantisation scale, $\beta$.

\subsection{Generalised Gamma Matrices} \label{Sec.5.3}

The construction of a full theory of quantum dynamics in smeared Minkoswki space, i.e., quantum field theory on a smeared space-time background, lies well beyond the scope of the present work. 
Nonetheless, the results of previous sections allow us to make limited conjectures about the description of relativistic electrons in such a theory. 
In particular, our previous results suggest that the kinetic term in the usual Dirac Equation (\ref{Dirac_eqn}) should be mapped according to

\begin{eqnarray} \label{}
i\hbar \gamma^{\mu}\partial_{\mu}\psi \rightarrow i\Gamma^{\mu}\mathcal{D}_{\mu}\Psi \, , 
\end{eqnarray}
where
\begin{eqnarray} \label{}
\mathcal{D}_{\mu} &:=& \hbar\partial_{\mu} + \beta\partial'_{\mu}
\nonumber\\
&:=& \hbar\frac{\partial}{\partial x^{\mu}}\bigg|_{(x'-x)^{\mu} = {\rm const.}} + \beta\frac{\partial}{\partial (x'-x)^{\mu}}\bigg|_{x^{\mu} = {\rm const.}} 
\end{eqnarray}
and
\begin{eqnarray} \label{gamma_matrices-mod}
\Gamma^{0} := 
\begin{bmatrix}
    0  &  \mathbb{1}_4 \\
    \mathbb{1}_4  & 0 
\end{bmatrix}
\, , \quad
\Gamma^{i} := \left(\frac{2}{\hbar + \beta}\right)
\begin{bmatrix}
    0  &  \hat{S}_{i}  \\
    -\hat{S}_{i}  &  0 
\end{bmatrix}
\, .
\end{eqnarray}

Less obvious is what happens to the canonical mass term, $-mc\psi$. 
In the momentum space of classical relativistic dynamics, the quantity analogous to the boost-invariant space-time interval, $s = \sqrt{t^2 - \vec{x}{\, }^2}$, is the invariant length of the $4$-momentum vector. 
Up to factors of $c^2$, this is simply the mass of the particle traversing the interval $s$: $mc^2 = \sqrt{E^2 - \vec{p}{\, }^2c^2}$. 

In canonical QFT, in which the space-time remains classical and sharply-defined, Lorentz invariance is preserved exactly. 
The classical mass appears as a parameter in the theory and is not promoted to the status of a quantum mechanical operator \cite{Peskin:1995ev}.
However, in a consistent theory of smeared space-time, we expect a radically different scenario. 
Intuitively, we would expect an appropriate smearing procedure to introduce an irremovable minimum uncertainty in the length of a space-time interval, $\Delta s$. 
Consistency of the position and momentum space pictures should then imply a corresponding minimum uncertainty in the length of the $4$-momentum vector, $\Delta m$. 
This is possible if the classical parameter $m$ is promoted to the status of a Hermitian operator, $m \mapsto \hat{m}$.

In \cite{Lake:2018zeg}, it was shown how to incorporate the effects of smearing directly into the definitions of observables. 
The resulting "smeared" Hermitian operators then act on the canonical quantum state $\ket{\psi} \in \mathcal{H}$. 
This formulation of the model yields exactly the same predictions as the smeared-state picture in which the fundamental state is $\ket{\Psi} \in \mathcal{H} \otimes \mathcal{H}$. 
However, in the smeared-operator picture, classical isometries are mapped to superpositions of isometries in the extended phase space of theory \cite{Lake:2018zeg}. 
So far, this method has only been applied to the translation generators of classical Euclidean space, but it may, in principle, be extended to the generators of other symmetries.  
Hence, we will address ways to implement smeared Lorentz symmetry, using this method, in a future publication. 
We hope that such an approach may be capable of yielding a natural definition of the mass operator $\hat{m}$.

\section{Discussion} \label{Sec.6}

\subsection{Conclusions} \label{Sec.6.1}

We have constructed generalised operators for angular momentum and spin in the smeared-space model of quantum geometry, originally proposed in \cite{Lake:2018zeg}. 
In this model, the canonical state $\ket{\psi} \in \mathcal{H}$ is mapped to the generalised "smeared" state, $\ket{\Psi} \in \mathcal{H} \otimes \mathcal{H}$. 
This represents the state of quantum matter, described by the wave function $\psi$, on a quantum background geometry. 
The latter is associated with an additional quantum state, $g$, so that $\Psi$ depends on both functions.  

In the original formulation of the smeared-space model, $\ket{\psi}$ and $\ket{g}$ are entangled, as proposed in the matter-geometry entanglement hypothesis \cite{Kay:2018mxr}. 
However, in Section \ref{Sec.4.2}, we defined a unitary operation that renders the smeared-state separable, yielding $\ket{\Psi} = \ket{\psi} \otimes \ket{g}$ (\ref{Psi_unitary_equiv}). 
The~transformation was inspired by the treatment of quantum reference frames (QRFs), considered in \cite{Giacomini:2017zju}, in which entanglement between subsystems of a composite state is frame-dependent. 

In the new basis, the generalised angular momentum operators can be written as the sum of three subcomponents, $\hat{L}_{i} = \hat{\mathcal{L}}_{i} + \hat{\mathcal{L}}'_{i} + \hat{\mathbb{L}}_{i}$ (\ref{L_sum*}). 
The first, $\hat{\mathcal{L}}_{i}$, which acts on the first subspace of the tensor product state, represents the angular momentum of a canonical quantum particle described by $\ket{\psi}$. 
The second, $\hat{\mathcal{L}}'_{i}$, which acts on the second subspace, represents the angular momentum of the quantum state associated with the background geometry, $\ket{g}$. 
The third subcomponent, $ \hat{\mathbb{L}}_{i}$, includes cross terms that act on both subspaces. 
This determines how quantum fluctuations of the background affect the angular momentum of particles propagating in the smeared geometry. 

The subcomponents $\left\{\hat{\mathcal{L}}_{i},\hat{\mathcal{L}}'_{i},\hat{\mathbb{L}}_{i}\right\}$ (\ref{L_sum*}) were found to obey a generalised algebra, defined by Equations (\ref{rearrange-L.1*})--(\ref{rearrange-L.5*}). 
These equations depend on two parameters, $\hbar$ and $\beta$, where the new parameter $\beta \simeq \hbar \times 10^{-61}$ is interpreted as the quantisation scale for geometry \cite{Lake:2018zeg}. 
Crucially, the generalised algebra implies the existence of GURs for angular momentum but recovers the canonical ${\rm so}(3)$ Lie algebra for the composite state of matter-plus-geometry, up to a simple rescaling $\hbar \rightarrow \hbar + \beta$. 
In~this respect, the angular momentum GURs are analogous to those for position and linear momentum, found in \cite{Lake:2018zeg}, in which the associated commutation relations are simply a rescaled representation of the Heisenberg algebra. 

Having constructed the generalised operators for orbital angular momentum, we considered the status of spin in the smeared background geometry. 
We argued, by analogy with the historical development of canonical QM, that the generalised spin operators should be finite-dimensional constant-valued matrices satisfying the same algebra as the components of angular momentum. 
Thus, we split the generalised spin operators into the sum of three terms, $\hat{S}_{i} = \hat{\mathcal{S}}_{i} + \hat{\mathcal{S}}'_{i} + \hat{\mathbb{S}}_{i}$ (\ref{S_sum}). 
By analogy with the subcomponents of $\hat{L}_{i}$ (\ref{L_sum*}), we required $\hat{\mathcal{S}}_{i}$ to act nontrivially on only the first spin-subspace of the tensor product smeared-state and $\hat{\mathcal{S}}_{i}$ to act nontrivially on only the second spin-subspace. 
The third subcomponent $\hat{\mathbb{S}}_{i}$, representing the interaction between the spin of the canonical quantum particle and the spin of the quantum state associated with the background geometry, was permitted to act nontrivially on both subspaces. 

We then required $\left\{\hat{\mathcal{S}}_{i},\hat{\mathcal{S}}'_{i},\hat{\mathbb{S}}_{i}\right\}$ to satisfy the algebra defined by Equations (\ref{rearrange-S.1})--(\ref{rearrange-S.5}), which are completely analogous to Equations (\ref{rearrange-L.1*})--(\ref{rearrange-L.5*}) under the interchange $\hat{\mathcal{L}}_{i} \leftrightarrow \hat{\mathcal{S}}_{i}$, $\hat{\mathcal{L}}'_{i} \leftrightarrow \hat{\mathcal{S}}'_{i}$, $\hat{\mathbb{L}}_{i} \leftrightarrow \hat{\mathbb{S}}_{i}$. 
We found that, assuming fermions with spin $\pm \hbar/2$ as the matter component of the composite state, 
this algebra can be satisfied {\it if and only if} the quantum state of the background has spin $\pm\beta/2$. 
Remarkably, therefore, consistency of the smeared-space spin algebra implies that the quantum state of the background space must be {\it fermionic} in nature. 
The implications of this result for the description of relativistic spin were briefly discussed in Section \ref{Sec.5.3}. 
Its possible implications for the physics of gravitions are discussed in Appendix \ref{Appendix.D}, where it was argued that these do not contradict existing results in quantum gravity theory.

Finally, the explicit forms of the generalised spin operators $\hat{S}_{x}$, $\hat{S}_{y}$ and $\hat{S}_{z}$ were also determined (\ref{S_i_explicit}). 
The composite smeared-background plus matter spin-state was found to have four eigenvectors, corresponding to two sets of $2$-fold degenerate eigenvalues, $\left\{3(\hbar+\beta)^2/4, \pm (\hbar+\beta)/2\right\}$. 
By analogy with the angular momentum case, the generalised spin algebra gives rise to GURs for spin measurements but recovers the canonical ${\rm su}(2)$ Lie algebra up to a simple rescaling, $\hbar \rightarrow \hbar + \beta$. 

\subsection{Future Work} \label{Sec.6.2}

To conclude, we consider the limitations of our present analysis and anticipate ways in which they may be overcome in future studies. 
Due to limitations of time and space, several key questions have not been addressed in the current work. 
These include the following: 

\begin{itemize}

\item We have not determined the spectral representations of the generalised angular momentum operators, $\left\{\hat{L}_i\right\}_{i=1}^{3}$, or the explicit form of their eigenstates. 
This is crucial because, without this spectrum, we are unable to determine how the re-smearing procedure, which forms part of the generalised measurement procedure in smeared-space (see Section \ref{Sec.3}), affects the form of the post-measurement states rendered by a measurement of $\hat{L}_i$. 
In \cite{Lake:2018zeg}, it was shown how re-smearing via the map (\ref{smear_map}) yields physical states as the outcomes of generalised position and momentum measurements. 
This also ensures that the minimum uncertainties, $\Delta_{\Psi} X^{i} \gtrsim l_{\rm Pl}$ and $\Delta_{\Psi} P_{j} \gtrsim m_{\rm dS}c$, hold for states {\it prepared} by such measurements. 
Thus, successive measurements can never violate these bounds. 
Na{\" i}vely, we would expect a similar result to hold for measurements of angular momentum, e.g., such that $\Delta_{\Psi} L_{i} \gtrsim  l_{\rm Pl}m_{\rm dS}c \simeq \beta$. 
This is in accordance with our intuition that perfectly sharp rotations cannot be performed on an unsharp background geometry. 

Furthermore, {\it if} such a fundamental limit to $\Delta_{\Psi} L_{i}$ exists due to re-smearing, it would be especially instructive to contrast this with our results for generalised spin measurements. 
In Section \ref{Sec.5.2}, the explicit forms of the generalised spin operators $\left\{\hat{S}_i\right\}_{i=1}^{3}$ were determined. 
Their eigenvectors and associated eigenvalues were also found and, in principle, we may use these to rewrite the spin-measurement operators in spectral form. 
However, in this case, there is no re-smearing procedure, since the ``smearing'' map Equation (\ref{smear_map}) applies only to the position-dependent part of the wave vector. 
Thus, eigenstates for which $\Delta_{\Psi} S_{i} = 0$ certainly exist. 
This is in accordance with our intuition that, as an {\it internal} property of the quantum particle, spin is not affected by the smearing of the external space in the same way as angular momentum. 
Unfortunately, in~the present work, we were not able to demonstrate the existence of a nonzero minimum bound on~$\Delta_{\Psi} L_{i}$. 

\item We did not consider multiparticle states. 
Hence, we did not attempt to generalise the Pauli exclusion principle (PEP) or the spin statistics theorem. 
This is a crucial and necessary step in the construction of a complete smeared-space generalisation of canonical QM. 
In particular, we note that the prediction of degenerate spin eigenstates, $\ket{\frac{3(\hbar + \beta)^2}{4}, \pm\frac{(\hbar + \beta)}{2}}$ and $\ket{\frac{3(\hbar + \beta)^2}{4}, \pm\frac{(\hbar + \beta)}{2}}_{\delta}$ (\ref{unentangled_eigenvectors})--(\ref{entangled_eigenvectors}), is potentially problematic for the model. 
For example, if the entangled and unentangled states in the spin ``up'' and spin ``down'' doublets are empirically indistinguishable, via measurements of $\hat{S}_z$ and $\hat{S}^2$, yet the spatial overlap of their associated wave functions is not forbidden by the generalised PEP, the model could be in immediate conflict with existing experimental data. 
That said, this may not be the case if the production of entangled states is extremely rare. 
This is not such an unreasonable assumption, since the interaction between the background and the canonical quantum fermions is characterised by a very small factor, $\sqrt{\hbar\beta} \simeq \hbar \times 10^{-30}$ (\ref{mathbb{S}_{i}}).   

\item We did not investigate, in detail, the potential consequences of our results for cosmology. 
In this respect, it is intriguing that consistency of the generalised spin algebra requires the quantum state associated with the background geometry to be fermionic. 
In \cite{Burikham:2015nma,Burikham:2017bkn,Lake:2017ync,Lake:2017uzd,Hashiba:2018hth}, it was shown how the pair-production of fermionic dark energy particles can generate the expansion of space {\it ad infinitum}. 
Remarkably, the particle mass required to generate the observed expansion rate is $m_{\Lambda} \simeq \sqrt{m_{\rm Pl}m_{\rm dS}} \simeq 10^{-3}$ eV. 
This is the unique mass scale that minimises the smeared-space GUR, Equation (\ref{GUR_X}). 
In this scenario, there exists a space-filling ``sea'' of dark energy fermions so that additional pair-production goes hand-in-hand with a concomitant production of {\it space}. 
This drives eternal universal expansion as the positive rest mass of the new particles is exactly cancelled by their negative gravitational energy (see \cite{Burikham:2015nma,Burikham:2017bkn,Lake:2017ync,Lake:2017uzd,Hashiba:2018hth} for details). 
Hence, it is clear that, {\it if} the fundamental quanta of space-time are fermionic, as suggested by the results obtained in the present work, universal expansion can also be viewed as a result of their continuous pair-production. 
Such a view is consistent with the model of particulate dark energy proposed in \cite{Burikham:2015nma,Burikham:2017bkn,Lake:2017ync,Lake:2017uzd,Hashiba:2018hth} and shares a number of qualitative features with the results of other studies. 
These include the model of space-time-matter (STM) ``atoms'', recently proposed in \cite{Singh:2019kep,Singh:2019hhi}. 

\end{itemize}

Finally we note that, given the close connection between the canonical angular momentum operators, the rotation generators in three-dimensional Euclidean space, and~the Lorentz generators in $(3+1)$-dimensional Minkowski space \cite{Rae,GroupTheoryJones,Peskin:1995ev}, the next logical step is to extend our analysis to the smearing of relativistic quantum field theories. 
This~should include "smeared" generalisations of the Maxwell, Klein--Gordon and Dirac equations, and, ultimately, of the QED Lagrangian. 
Clearly, many conceptual and mathematical problems must be resolved before smeared-space QFTs can be rigorously defined, but the results presented herein represent a first step towards their construction.

\begin{center}
{\bf Acknowledgments}
\end{center}

SDL was supported by the Natural Science Foundation of Guangdong Province, grant no. 2016A030313313.


\appendix

\section{Subtleties with Angular Momentum in Classical Mechanics and Canonical QM} \label{Appendix.A}

In this appendix, we consider a number of subtleties that arise in the canonical treatment of angular momentum for classical point-particles in flat Euclidean space. 
These, in turn, have implications for the treatment of quantum particles in flat space (that is, for canonical QM) and, hence, for any would-be theory of quantum geometry in which degrees of freedom are associated with the spatial background.

\subsection{Classical Mechanics} \label{Appendix.A.1}

In classical mechanics, the angular momentum pseudo-vector of a point-particle in three-dimensional Euclidean space is
\begin{eqnarray} \label{classical_am}
\vec{l} = \vec{r} \times \vec{p} \, ,
\end{eqnarray}
where $\vec{r}$ is the position vector, relative to the origin of rotation, and~$\vec{p}$ is the instantaneous linear momentum. 
The cross denotes the vector product which, for an arbitrary pair of vectors, is defined as
$\vec{a} \times \vec{b} = |\vec{a}| |\vec{b}| \sin\theta \, \underline{\bold{n}}$, where $\underline{\bold{n}}$ is the unit vector perpendicular to the plane defined by $\vec{a}$ and $\vec{b}$ and $\theta$ is the angle between them. 

Strictly, both the vector and scalar products are defined between pairs of vectors at the same spatial point. 
Thus, we must parallel transport the linear momentum vector to the origin of the displacement vector, since, by convention, the angular momentum vector is defined at the centre of rotation \cite{LandauMechanics}. 
More formally, if the position of the particle ``$x$'' is specified by the coordinates $\left\{x^{i}\right\}_{i=1}^{3}$, 
the true linear momentum is given by $\vec{p}(x) = p_{i}(x)\underline{\bold{e}}^i(x)$,
where $\left\{\underline{\bold{e}}_i(x)\right\}_{i=1}^{3}$ $\left(\left\{\underline{\bold{e}}^i(x)\right\}_{i=1}^{3}\right)$ span the tangent (cotangent) space at $x$. 
(Here, we do not explicitly include the time parameter in the argument of the vector components for the sake of notational elegance, i.e., we use $p_i(x)$ rather than $p_i(x,t)$. 
This convention is followed throughout the remainder of the text, e.g., $q^{i}(x)$ and $\dot{q}^{i}(x)$, where a dot denotes differentiation with respect to $t$ and $x$ denotes the spatial coordinates.) 
The vector ``$\vec{p}\,$'' appearing in Equation (\ref{classical_am}) is then $\vec{p}(0) = p_{i}(0)\underline{\bold{e}}^i(0)$, where $p_{i}(0) = \Gamma^{x_2}_{x_1}(\gamma) p_{i}(x)$, and~the operator $\Gamma^{x_2}_{x_1}(\gamma)$ parallel transports vectors from $\gamma(x_1)$ to $\gamma(x_2)$ along the curve $\gamma$ \cite{Frankel:1997ec}. 

Similar considerations hold when we take the scalar product, $\braket{\vec{a}(x),\vec{b}(x)} = g_{ij}(x)a^{i}(x)b^{j}(x) = a_{i}(x)b^{i}(x)$, which is often written simply as $\vec{a}.\vec{b} = |\vec{a}| |\vec{b}| \cos\theta$ (that is, without specifying the point $x$ at which it is defined) for the sake of notational simplicity. 
For example, the $\vec{p}$ appearing in the usual dot product $\vec{p}.\vec{r}$, where $\vec{r}$ is the displacement vector, is in fact the parallel-transported vector $\vec{p}(0) = p_{i}(0)\underline{\bold{e}}^i(0)$, considered above.

In Euclidean space, parallel transport is path-independent and also preserves the inner product, i.e., $\braket{\Gamma(\gamma)\vec{a},\Gamma(\gamma)\vec{b}}(x_1) = \braket{\vec{a},\vec{b}}(x_2)$ for arbitrary start and end points, $x_1$ and $x_2$, on any path $\gamma$ \cite{Frankel:1997ec}. 
It follows that the vector product between any pair of vectors is also preserved. 
Thus, location-independent meanings can be ascribed to the quantities ``$\vec{a}.\vec{b}$'' and ``$\vec{a} \times \vec{b}$'', which justifies the usual neglect of such subtleties for systems defined in a Euclidean background.  

However, in curved geometries such nice properties do not, in general, hold. 
In fact, notions such as vector displacement can be only defined locally, e.g., via $d\vec{r} = dx^{i}\underline{\bold{e}}_{i}(x)$. 
Thus, expressions involving a finite-length displacement vector ``$\vec{r}\,$'', such as ``$\vec{p}.\vec{r}\,$'' and ``$\vec{r} \times \vec{p}\,$'', do not make sense. 
The integral of $d\vec{r}$ from $x_1$ to $x_2$ is path-dependent, and~the result is not a genuine vector, so that curved geometries are not vector spaces \cite{Nakahara:2003nw,Frankel:1997ec}. 
Similarly, the canonical momentum $\vec{p}$ may be seen as a displacement vector in Euclidean momentum space, whereas only the local quantity $d\vec{p} = dp_{i}(x)\underline{\bold{e}}^{i}(x)$ can be consistently defined for particles in curved backgrounds. 

We recall that the phase space of a classical system is given by the cotangent bundle of the manifold on which the system is defined. 
The classical metric then gives an isomorphism between the tangent and cotangent bundles \cite{Frankel:1997ec}. 
Hence, both physical space and momentum space may be curved. 
In this case, angular momentum also becomes a local property and is conserved only locally as a result of local (not global) rotational symmetry. 

We also recall that, for an arbitrary Riemannian geometry, the inner product between tangent vectors defines the metric,
\begin{eqnarray} \label{classical_metric}
g_{ij}(x) = \braket{\underline{\bold{e}}_i(x),\underline{\bold{e}}_j(x)} \, .
\end{eqnarray}
In Euclidean geometry, the metric takes a particularly simple form {\it in Cartesian coordinates}, i.e., $\eta_{ij}(x) = {\rm diag}(1,1,1)$ for all $x$. 
Thus, in Cartesian coordinate systems, there is no distinction between contravariant and covariant components, $dx^{i} = dx_{i}$, or between the tangent and cotangent vectors, $\underline{\bold{e}}_{i}(x) = \underline{\bold{e}}^{i}(x)$. 
Such coordinate systems are extremely special.
In particular, they are the only {\it globally orthogonal} coordinates that exist, even in flat Euclidean space.

By ``globally orthogonal'' we mean that (i) any two orthogonal (co)tangent vectors will remain orthogonal if parallel transported along different paths to any other point, and~(ii) the parallel-transported vectors will be equal to the corresponding (co)tangent vectors defined at the new point. 
Note that, although the first condition holds for locally orthogonal coordinate systems in flat space, e.g., cylindrical and spherical polars, the second condition does not, since the lengths of $\underline{\bold{e}}_{i}(x)$ and / or $\underline{\bold{e}}^{i}(x)$ vary from point to point.
In non-Euclidean spaces, there are no such globally orthogonal coordinate systems \cite{Frankel:1997ec}.
 
Cartesian coordinates are, therefore, the only coordinates that give rise to a set of {\it globally orthonormal} tangent vectors.
In three dimensions, these are defined by the algebra:
\begin{eqnarray} \label{orthonormal_basis_3D}
\underline{\bold{e}}_i(x) \times \underline{\bold{e}}_j(x) = \epsilon_{ij}{^{k}}\underline{\bold{e}}_k(x) \, ,
\end{eqnarray}
$i,j,k \in \left\{1,2,3\right\}$, which holds for the tangent vectors defined at all points $x$. 
Here, $\epsilon_{ij}{^{k}}$ is the Levi-Civita symbol.  
This is defined as $\epsilon_{ij}{^{k}} = 1$ for cyclic permutations of $ijk$, $\epsilon_{ij}{^{k}} = -1$ for non-cyclic permutations, and~$\epsilon_{ij}{^{k}} = 0$ otherwise, but it is not a tensor. 
Put simply, the algebra (\ref{orthonormal_basis_3D}) is defined in terms of a symbol, $\epsilon_{ij}{^{k}}$, rather than a tensor, because it is valid only in a specific coordinate system.

Hence, even in three-dimensional Euclidean space, the only set of basis vectors satisfying Equation~(\ref{orthonormal_basis_3D}) are tangent to the''curves'' (i.e., straight lines) defined by the conditions $x^{i} = {\rm const.}$, where $\left\{x^{i}\right\}_{i=1}^{3}$ are the Cartesian coordinates $\left\{x,y,z\right\}$. 
In this case, the relevant Poisson brackets (PB)~are
\begin{eqnarray} \label{PB_xx_pp}
\left\{x^{i},x^{j}\right\}_{\rm PB} = 0 \, , \quad \left\{p_{i},p_{j}\right\}_{\rm PB} = 0 \, , 
\end{eqnarray}
\begin{eqnarray} \label{PB_xp}
\left\{x^{i},p_{j}\right\}_{\rm PB} = \delta^{i}{}_{j} \, ,  
\end{eqnarray}
and
\begin{eqnarray} \label{PB_lx_lp}
\left\{l_{i},x^{k}\right\}_{\rm PB} = \epsilon_{ij}{^{k}} x^{j} \, , \quad \left\{l_{i},p_{j}\right\}_{\rm PB} = \epsilon_{ij}{^{k}} p_{k} \, ,
\end{eqnarray}
\begin{eqnarray} \label{PB_ll}
\left\{l_{i},l_{j}\right\}_{\rm PB} = \epsilon_{ij}{^{k}}l_{k} \, ,  
\end{eqnarray}
\begin{eqnarray} \label{PB_l^2l}
\left\{l^2,l_{i}\right\}_{\rm PB} = 0 \, ,  
\end{eqnarray}
where the components of the angular momentum vector $\vec{l} := l_{i}(0)\underline{\bold{e}}^i(0)$ are given by
\begin{eqnarray} \label{l_i}
l_{i} = (\vec{r} \times \vec{p})_{i} = \epsilon_{ij}{^{k}}x^{j}p_{k} \, .
\end{eqnarray} 

The structures of the canonical Poisson brackets for the components of the position and linear momentum vectors (\ref{PB_xx_pp})--(\ref{PB_xp}), as well as for the components of the angular momentum (\ref{PB_lx_lp})--(\ref{PB_l^2l}), are therefore intimately related to both the geometric structure of Euclidean space and, crucially, to the specific choice of coordinates used to describe physical systems (\ref{orthonormal_basis_3D}). 
More generally, the components of the angular momentum along arbitrary vector directions, $\vec{a}$ and $\vec{b}$, are related via 
\begin{eqnarray} \label{PB_ll*}
\left\{l_{\vec{a}},l_{\vec{b}}\right\}_{\rm PB} = \sin\theta \, l_{\underline{\bold{n}}} \, ,
\end{eqnarray}
where $l_{\underline{\bold{n}}} = \braket{\underline{\bold{n}},\vec{l}}$, etc. 
Equation (\ref{PB_ll*}) holds for any $\vec{a}$ and $\vec{b}$ in flat Euclidean space and is independent of the choice of coordinates. Equation (\ref{PB_ll}) is then recovered by choosing a global Cartesian basis. 

It is important to note that, although analogues of Equations (\ref{PB_xx_pp})--(\ref{PB_xp}) exist for any set of canonically conjugate phase space coordinates, $\left\{q^{i},\pi_{j}\right\}$, analogues of Equations (\ref{PB_lx_lp})--(\ref{l_i}) do not. 
Specifically, any set of generalised position coordinates, 
\begin{eqnarray} 
q^{i} = q^{i}(x) \, , 
\end{eqnarray}
together with their canonically conjugate momenta,
\begin{eqnarray} 
\pi_{j} = \frac{\partial L}{\partial \dot{q}^{j}} \, , 
\end{eqnarray}
where $L(q,\dot{q})$ is the Lagrangian of the classical system, satisfy
\begin{eqnarray} \label{PB_qq_pipi}
\left\{q^{i},q^{j}\right\}_{\rm PB} = 0 \, , \quad \left\{\pi_{i},\pi_{j}\right\}_{\rm PB} = 0 \, , 
\end{eqnarray}
and
\begin{eqnarray} \label{PB_qpi}
\left\{q^{i},\pi_{j}\right\}_{\rm PB} = \delta^{i}{}_{j} \, .
\end{eqnarray}
The bracket structure (\ref{PB_qq_pipi})--(\ref{PB_qpi}) is then preserved by any canonical coordinate transformation \cite{LandauMechanics}. 

When $q^{i} = x^{i}$ are Cartesian coordinates, $\pi_{j} = p_{j}$ are the components of the physical linear momentum, and~the corresponding components of the angular momentum are given by Equation (\ref{l_i}).
However, under a general canonical transformation, $x^{i} \rightarrow q^{i}$, $p_{j} \rightarrow \pi_{j}$, where the new phase space coordinates are not Cartesian, the transformed components of the physical angular momentum are not given by a formula analogous to (\ref{l_i}). 
Although we may define the analogous quantities $\xi_{i} := \epsilon_{ij}{}^{k}q^{j}\pi_{k}$, these do not, in general, correspond to components of the angular momentum vector, unless $\left\{q^{i},\pi_{j}\right\}$ represent Cartesian phase space coordinates.
Similarly, it is straightforward to show that $\left\{\xi_{i}\right\}_{i=1}^{3}$ do not satisfy the algebras (\ref{PB_lx_lp})--(\ref{PB_l^2l}), unless $q^{i} = x^{i}$, $\pi_{j} = p_{j}$ ($\xi_{i} = l_{i}$).

In non-canonical coordinate systems, the canonical Poisson bracket structures (\ref{PB_qq_pipi})--(\ref{PB_qpi}) are also destroyed \cite{LandauMechanics}. 
Nonethless, in general curved spaces, non-Cartesian canonical coordinates can always be defined (at least locally \cite{Frankel:1997ec}) so that Equations (\ref{PB_qq_pipi})--(\ref{PB_qpi}) may still be satisfied for an appropriate choice of $\left\{q^{i}, \pi_{j}\right\}$. 
Despite this, the physical space displacement vector ``$\vec{r}\,$'' and momentum space displacement vector ``$\vec{p}\,$'' are not well defined for systems in curved geometries, so that Equation (\ref{l_i}) does {\it not} hold, regardless of our choice of coordinates. 
In summary, the relation (\ref{l_i}) is extremely special. It holds only in flat Euclidean space and in Cartesian phase space coordinates $\left\{x^i,p_j\right\}$.
 
\subsection{Canonical QM} \label{Appendix.A.2}

In quantum mechanics, canonical quantisation is always performed in Cartesian coordinates, $\left\{x^i,p_j\right\}$ \cite{Messiah:book}. 
Specifically, one obtains the operators corresponding to the classical values $x^{i}$ and $p_{j}$ by performing the map:
\begin{eqnarray} \label{map-1}
x^i \mapsto \hat{x}^i \, , \quad p_i \mapsto \hat{p}_i \, , 
\end{eqnarray}
where 
\begin{eqnarray} \label{x_op}
\hat{x}^{i} = \int x^{i} \ket{\vec{r}}\bra{\vec{r}} {\rm d}^{3}\vec{r} \, , \quad \hat{p}_{j} = \int p_{j} \ket{\vec{p}}\bra{\vec{p}} {\rm d}^{3}\vec{p} \, , 
\end{eqnarray}
and 
\begin{eqnarray} \label{deBroglie}
\braket{\vec{r} | \vec{p}} = \frac{1}{\sqrt{2\pi \hbar}} \exp\left(\frac{i}{\hbar} \vec{p}.\vec{r}\right) \, . 
\end{eqnarray}
Here, we use the shorthand ${\rm d}^{3}\vec{r} = \sqrt{\det g_{ij}(x)}{\rm d}x^3$ and ${\rm d}^{3}\vec{p} = \sqrt{\det \tilde{g}_{ij}(p)}{\rm d}^3p$, where $g_{ij}(x)$ and $\tilde{g}_{ij}(p)$ denote the metrics on the position and momentum space submanifolds of the classical phase space, respectively. 
In Cartesians, ${\rm d}^{3}\vec{r} = {\rm d}x^3$ and ${\rm d}^{3}\vec{p} = {\rm d}^3p$, but we may perform the integration in any coordinates we wish as long as $x^{i}$ and $p_{j}$ represent Cartesian components of $\vec{x}$ and $\vec{p}$.

Equation (\ref{deBroglie}) is equivalent to the canonical de Broglie relation between momentum and wave number, $\vec{p} = \hbar \vec{k}$, which, together with Equation (\ref{x_op}), yields
\begin{eqnarray} \label{xx_pp_comm}
[\hat{x}^{i},\hat{x}^{j}] = 0 \, , \quad [\hat{p}_{i},\hat{p}_{j}] = 0 \, , 
\end{eqnarray}
and
\begin{eqnarray} \label{Heisenberg}
[\hat{x}^{i},\hat{p}_{j}] = i\hbar \delta^{i}{}_{j} \hat{\mathbb{1}} \, .
\end{eqnarray}
The canonical commutators (\ref{xx_pp_comm})--(\ref{Heisenberg}) are the quantum counterparts of the classical Poisson brackets (\ref{PB_xx_pp})--(\ref{PB_xp}) and are consistent with the general correspondence \cite{DiracQM:book}
\begin{eqnarray} \label{correspondence}
\left\{O_1,O_2\right\}_{\rm PB} = \lim_{\hbar \rightarrow 0} \frac{1}{i\hbar} [\hat{O}_1,\hat{O}_2] \, ,
\end{eqnarray}
where $O(x,p)$ is a function of the Cartesian coordinates of the classical phase space and $\hat{O}(\hat{x},\hat{p})$ has the same functional form with respect to the corresponding operators (assuming resolution of any potential ordering problems). 

However, we note that, in order to interpret the results of physical measurements, coordinate values alone are not enough: one must also know where in physical space the coordinate ``axes'' are located. 
For example, in order to reach the point in space labelled by the coordinates $\left\{x^{i}\right\}_{i=1}^{3}$, one must begin at the origin and travel $x^{i}$ units along the $i^{\rm th}$ coordinate direction, keeping all other coordinate values fixed, for each $i$ sequentially. 
This procedure is general and holds regardless of whether each line $x^{i} = {\rm const.}$ defines a linear or a curvilinear ``axis''. 

Thus, operationally we may say that, in order to detect a particle at ``$x$'', one must receive a signal emitted from the physical point defined by both the coordinates $\left\{x^{i}\right\}_{i=1}^{3}$ and the associated coordinate axes. 
Furthermore, since the tangent vectors $\left\{\underline{{\bf e}}_{i}(x)\right\}_{i=1}^{3}$ are tangent to the curves $x^{i} = {\rm const.}$ (in any coordinate system), it is clear that knowledge of the metric (\ref{classical_metric}) is required in order to interpret coordinate values as positions in physical space.

This is the case in classical mechanics and remains the case in canonical QM, in which the background space is assumed to be fixed and classical. 
Specifically, in three-dimensional Euclidean space, the displacement vector and (parallel-transported) momentum vector of a classical particle may be written as
\begin{eqnarray} \label{vecs}
\vec{r} = x^i\underline{\bold{e}}_i(0) \, , \quad \vec{p} = p_j\underline{\bold{e}}^j(0) \, ,
\end{eqnarray}
respectively, where $\underline{\bold{e}}_i(0)$ denote Cartesian tangent vectors defined at the origin. 
Quantising the system by ``promoting'' coordinates but {\it not} tangent vectors to operators is equivalent to quantising matter (i.e., point-particles) while leaving the background geometry, which is defined by the classical metric (\ref{classical_metric}), unchanged. 
Practically, this implies the {\it de facto} definition of a map:
\begin{eqnarray} \label{map-1}
\vec{r} \mapsto \hat{\vec{r}} \, , \quad \vec{p} \mapsto \hat{\vec{p}} \, , 
\end{eqnarray} 
(note the hat above the vector arrow), where 
\begin{eqnarray} \label{map-2}
\hat{\vec{r}} := \hat{x}^i\underline{\bold{e}}_i(0) \, , \quad \hat{\vec{p}} := \hat{p}_j\underline{\bold{e}}^j(0) \, .
\end{eqnarray}

Though such a definition does not form part of the abstract Hilbert space formalism of canonical QM, in which the spectral representation of the Hermitian operators (\ref{x_op}) is agnostic to their physical interpretation \cite{Ish95}, it is undoubtedly necessary in order for real-world experimentalists to connect the predictions of this formalism with the outcomes of real-world measurements. 
This implies a subtle link between the mathematical structure of quantum systems (abstract Hilbert spaces) and the mathematical structure of physical spaces (symplectic manifolds and their associated geometries), which is of vital importance for the problem of quantum gravity. 

The existence of this link is especially highlighted when one considers the quantisation of angular momentum. 
Since the classical formula (\ref{classical_am}) is a relation between vector quantities, one would expect the tangent vector part of this expression to be affected by the quantisation of geometry, just as the component part is affected by the quantisation of matter. 
In any would-be theory of quantum matter living ``in'' a quantum geometry, both aspects must be accounted for. 

In relativistic quantum gravity, quantum fluctuations of the background geometry are expected to induce curvature fluctuations and hence fluctuations in the gravitational field strength over very small length scales comparable to the Planck length, $l_{\rm Pl} \simeq 10^{-33}$ cm. 
\footnote{This remains true even if the effects of classical space-time curvature can be neglected over such small intervals. In~Planck-sized volumes $\sim l_{\rm Pl}^4$, the classical background space-time may be regarded as approximately flat as long as its curvature remains significantly below the Planck curvature, $\sqrt{K} \sim 1/l_{\rm Pl}^2$, where $K$ is the Kretschmann scalar. Thus, classical curvature is typically negligible in such regions, except in extreme scenarios, such as close to the singularity of a black hole \cite{Chandrasekhar:book}.} 
This, in turn, is expected to give rise to a minimum resolvable length scale of the order of $l_{\rm Pl}$ \cite{Crowell:2005ax}. 
Assuming that fluctuations in the background space-time include fluctuations in the space-space part of the metric, $g_{ij}$, a non-zero spatial (Riemannian) curvature, $R_{ijkl} \neq 0$, is also generated, in addition to the nonzero space-time (pseudo-Riemannian) curvature, $R_{\mu\nu\rho\sigma} \neq 0$.
In this scenario, such fluctuations give rise to two effects that are relevant to our present discussion. 

First, they destroy the Cartesian coordinate system on which the canonical quantisation {\it of coordinates} is based. 
(We recall, again, that global Cartesian coordinates do not exist in curved space.)
Second, they destroy the physical rationale for the quantisation of coordinates {\it alone}, while leaving the tangent vectors with which they are associated, i.e., the coordinate axes and the geometry they define via the classical metric (\ref{classical_metric}), unchanged.
Equivalently, we may say that they destroy the coordinate system on which the map that defines the canonical quantisation of {\it matter}, (\ref{map-1})--(\ref{map-2}), is based. 
In addition, their very existence implies the need to define a new map, between the classical tangent vectors $\left\{{\bf \underline{e}}_{i}\right\}_{i=1}^{3}$ and a new set of operators $\left\{\hat{{\bf \underline{e}}}_{i}\right\}_{i=1}^{3}$, which represents the quantisation of the background {\it geometry} in which the quantum matter lives. 

However, even in the study of quantum gravity phenomenology, such subtleties can easily be neglected, {\it if} we restrict our attention to the quantum counterparts of classical relations involving only coordinates. 
This is true for all studies of modified position-momentum commutators, in which one assumes the usual correspondence $\left\{x^{i},p_{j}\right\}_{\rm PB} = \lim_{\hbar \rightarrow 0} \frac{1}{i\hbar} [\hat{x}^{i},\hat{p}_{j}]$ but modifies either the right-hand side, such that $\frac{1}{i\hbar} [\hat{x}^{i},\hat{p}_{j}] \neq \delta^{i}{}_{j} \, \hat{\mathbb{1}}$, or both the left- and right-hand sides simultaneously \cite{Kempf:1994su,Hossenfelder:2014ifa}. It is also true of recent studies of the smeared-space model in which relations between vector quantities were similarly neglected \cite{LakeUkraine2019,Lake:2018zeg}. 

Nonetheless, it is clear that such subtleties cannot be neglected when one explicitly considers the counterparts of vector relations, such as Equation (\ref{classical_am}), on a quantum background. 
In this case, the coordinate-dependent expression $l_{i} = \epsilon_{ij}{^{k}}x^{j}p_{k}$ (\ref{l_i}) and its associated algebra (\ref{PB_ll})--(\ref{PB_l^2l}) emerge only after taking the inner product of (\ref{classical_am}) with the relevant tangent vector: $l_{i} = \braket{\underline{\bold{e}}_{i}(0),\vec{l}} = l_{j}\braket{\underline{\bold{e}}_{i}(0),\underline{\bold{e}}^{j}(0)} = l_{j}\delta^{j}{}_{i}$. 
Clearly, this relies on the definition of the classical metric (\ref{classical_metric}). 

In canonical QM, this poses no problems, since we are not required to quantise the background space.
Thus, defining $\hat{\vec{r}}$ and $\hat{\vec{p}}$ via Equations (\ref{map-1})--(\ref{map-2}), we may define the vector angular momentum operator as 
\begin{eqnarray} \label{QM_am}
\hat{\vec{l}} = \hat{\vec{r}} \times \hat{\vec{p}} \, .
\end{eqnarray}
It follows immediately that
\begin{eqnarray} \label{commutators_lx_lp}
[\hat{l}_{i},\hat{x}^{k}] = i\hbar \, \epsilon_{ij}{}^{k} \hat{x}^{j} \, , \quad 
[\hat{l}_{i},\hat{p}_{j}] = i\hbar \, \epsilon_{ij}{}^{k} \hat{p}_{k} \, ,
\end{eqnarray}
and
\begin{eqnarray} \label{ll_commutator}
[\hat{l}_{i},\hat{l}_{j}] = i\hbar \, \epsilon_{ij}{^{k}}\hat{l}_{k} \, ,  
\end{eqnarray}
\begin{eqnarray} \label{l^2l_commutator}
[\hat{l}^2,\hat{l}_{i}] = 0 \, ,  
\end{eqnarray}
where the Cartesian components of $\hat{\vec{l}} := \hat{l}_{i}\underline{\bold{e}}^i(0)$ are given by
\begin{eqnarray} \label{l_i_QM}
\hat{l}_{i} = (\hat{\vec{r}} \times \hat{\vec{p}})_{i} = \epsilon_{ij}{}^{k}\hat{x}^{j}\hat{p}_{k} \, .
\end{eqnarray}
These are obtained as $\hat{l}_{i} = \braket{\underline{\bold{e}}_{i}(0),\hat{\vec{l}}} = \hat{l}_{j}\braket{\underline{\bold{e}}_{i}(0),\underline{\bold{e}}^{j}(0)} = \hat{l}_{j}\delta^{j}{}_{i}$, by complete analogy with the classical case so that
\begin{eqnarray} \label{l_vec_commutator}
[\hat{l}_{\vec{a}},\hat{l}_{\vec{b}}] = i\hbar \, \sin\theta \hat{l}_{\underline{\bold{n}}} \, . 
\end{eqnarray}

By the Schr{\" o}dinger--Robertson relation, Equations (\ref{ll_commutator}) and (\ref{l_vec_commutator}) then give rise to the uncertainty relations
\begin{eqnarray} \label{ll_UR}
\Delta_{\psi}l_{i}\Delta_{\psi}l_{j} \geq \frac{\hbar}{2} |\epsilon_{ij}{^{k}}\braket{\hat{l}_{k}}_{\psi}| \, ,  
\end{eqnarray}
\begin{eqnarray} \label{l_vec_UR}
\Delta_{\psi}l_{\vec{a}}\Delta_{\psi}l_{\vec{b}} \geq \frac{\hbar}{2} |\sin\theta \braket{\hat{l}_{\underline{\bold{n}}}}_{\psi}| \, ,
\end{eqnarray}
respectively. 
Equation (\ref{l_vec_UR}) is the more general relation, and~the uncertainty relation for Cartesian components is recovered by taking $\theta = m(\pi/2)$ ($m \in \mathbb{Z}$). 
Similarly, Equations (\ref{commutators_lx_lp}) give rise to
\begin{eqnarray} \label{UR_lx_lp}
\Delta_{\psi}l_{i}\Delta_{\psi}x^{k} &\geq& \frac{\hbar}{2} \, |\epsilon_{ij}{}^{k} \braket{\hat{x}^{j}}_{\psi}| \, , 
\nonumber\\
\Delta_{\psi}l_{i}\Delta_{\psi}p_{j} &\geq& \frac{\hbar}{2}\, |\epsilon_{ij}{}^{k} \braket{\hat{p}_{k}}_{\psi}| \, .
\end{eqnarray}

We stress that in canonical QM one quantises $\vec{r}$, $\vec{p}$ and $\vec{l}$ by quantising the relevant vector components, $x^{i}$, $p_{i}$ and $l_{i}$, respectively, but leaving the associated classical basis vectors $\left\{\underline{\bold{e}}_{i}(x)\right\}_{i=1}^{3}|_{x=0}$ unchanged. 
This is a subtle mathematical expression of the fact that canonical quantum systems are described by superpositions of eigenstates (e.g., position, linear momentum, or angular momentum eigenstates) that live on, or ``in'', a fixed classical background. 
More concretely, we may say that canonical quantum wave functions $\psi(\vec{r})$ are defined as complex-valued fields on the metric space defined by the tangent vectors $\left\{\underline{\bold{e}}_{i}(x)\right\}_{i=1}^{3}$ in Equation (\ref{classical_metric}). 

In a true quantum gravity scenario, in which the background itself is subject to quantum fluctuations associated with the minimum length scale \cite{Crowell:2005ax}, this picture must be radically revised. 
In this paper, however, we have considered one scenario, which may be regarded as a more conservative solution to this problem. 
In Section \ref{Sec.2}, we outlined an alternative interpretation of the smearing function $g(\vec{r}{\, '} - \vec{r})$, introduced in the original formulation of the smeared-space model \cite{Lake:2018zeg}, in which quantum fluctuations of the background give rise to a minimum length but {\it not} to curvature fluctuations in the non-relativistic limit. 
The new interpretation and its physical consequences are discussed in greater detail in Appendix \ref{Appendix.B}. 

Though the resulting model does not include gravity-as-curvature, we argue that it is consistent with existence of the Newtonian gravitational potential, viewed, in the standard way, as a scalar field in flat Euclidean space \cite{LandauMechanics,DiracGR:book}. (See Appendix \ref{Appendix.B}.) 
Thus, our quantisation procedure may be implemented via a map between classical coordinates and Hermitian operators, as in canonical QM. 
Furthermore, it is implemented as a map between {\it Cartesian coordinates} in the classical phase space and generalisations of the canonical operators $\hat{x}^{i}$ and $\hat{p}_{j}$. 
This justifies the assumptions made in Section \ref{Sec.4.1}, when defining the generalised {\it vector} operators $\hat{\vec{R}}$, $\hat{\vec{P}}$ and $\hat{\vec{L}}$ by analogy with the canonical theory.
The generalised coordinate operators, denoted $\hat{X}^{i}$ and $\hat{P}_{j}$, act on a composite quantum state incorporating both matter and geometry. 
In our model, points in the quantum background are subject to stochastic movements, but these do not change the underlying {\it flat} geometry of the space. 


\section{Physical Interpretation of the Smearing Function, Revisited} \label{Appendix.B}

In \cite{Lake:2018zeg}, the smearing function $g(\vec{r}{\, '}-\vec{r})$ was interpreted as the probability amplitude for the transition $\vec{r}{\, '} \rightarrow \vec{r}$. 
Importantly, this allows (at least in principle) for the smeared-space model to describe curvature fluctuations in the background geometry. 
The mechanism for this is as follows. 
Since, in this formulation of the model, only primed variables are physically accessible, an arbitrary set of measured values $(\vec{r}{\, '})$ determines a three-dimensional submanifold in the extended six-dimensional phase space $(\vec{r},\vec{r}{\, '})$. 
This may be  described by an arbitrary vector function, $\vec{r}{\, '}(\vec{r})$. 
Hence, if the metric on the $(\vec{r},\vec{r}{\, '})$ hyperplane is known, a natural choice for the metric on the $\vec{r}{\, '}(\vec{r})$ submanifold is the induced metric, which is obtained by performing the push-forward from the metric in the six-dimensional bulk space \cite{Nakahara:2003nw,Frankel:1997ec}. 

In the original analysis of the smeared-space model \cite{Lake:2018zeg}, it was argued that consistency requires the coordinates $(\vec{r},\vec{r}{\, '})$ to label points in a flat pseudo-Riemannian manifold with $3$ space-like dimensions and $3$ time-like dimensions, i.e., a $(3+3)$-dimensional generalised Minkowski space with metric signature $(+++---)$. 
Despite this, however, induced metrics on observable three-dimensional submanifolds have positive signature, $(+++)$, so that the model describes non-relativistic matter on a fluctuating spatial background. 
(See \cite{Lake:2018zeg} for details.) 
In principle, these fluctuations can give rise to arbitrary Riemannian geometries, but practically, the probability amplitudes for transitions with $|\vec{r}{\, '}-\vec{r}| \gg l_{\rm Pl}$ are vanishingly small. 
Indeed, if $|g(\vec{r}{\, '}-\vec{r})|$ is peaked at $\vec{r}{\, '} = \vec{r}$, as is the case for Gaussian smearing, the most probable geometry is isomorphic to the original, flat, Euclidean space. 
Transitions within one standard deviation of $|g(\vec{r}{\, '}-\vec{r})|$ remain relatively likely, but these correspond to small fluctuations of order $|\vec{r}{\, '}-\vec{r}| \lesssim l_{\rm Pl}$, as expected phenomenologically \cite{Crowell:2005ax}. 

In Appendix \ref{Appendix.A}, we discussed the subtle ways in which canonical quantisation techniques encode assumptions about the nature of the background geometry in which material systems ``live''. 
In~particular, we explained why the standard procedure of promoting classical coordinates to Hermitian operators is not applicable in the presence of spatial curvature. 
This is especially obvious for physical quantities that depend on the canonical displacement vectors, $\vec{r}$ and $\vec{p}$, which are only well defined in Euclidean geometries, and, hence, is especially problematic for the quantisation of angular momentum in curved space. 

To overcome this problem, we took a stricter interpretation of the smearing function in the present work. 
Instead of allowing arbitrary transitions $\vec{r} \rightarrow \vec{r}{\, '}$ in the extended $(\vec{r},\vec{r}{\, '})$ phase space, we limited the available transitions to the {\it pair-wise exchange} of points. 
Thus, we interpret $g(\vec{r}{\, '}-\vec{r})$ as the probability amplitude for the transition $\vec{r} \leftrightarrow \vec{r}{\, '}$. 
This is a far more restrictive condition, but it is straightforward to verify that the quantitative results presented in Section \ref{Sec.3} are unaffected by our interpretation of $g$. 
Nonetheless, the new interpretation has several advantages.

First, since Euclidean spaces of any dimension are maximally symmetric, all points are considered equivalent. 
It follows immediately that the pair-wise exchange of points, or even of whole neighbourhoods surrounding $\vec{r}$ and $\vec{r}{\, '}$ \cite{Isham:1999qu}, cannot change the geometry of the underlying space. 
Thus, ``quantising'' Euclidean geometry in this way introduces an additional stochastic variation into the quantum measurement procedure, but the resulting fluctuations cannot generate spatial curvature. 
With this interpretation, we are still able to generate GURs and to derive the dark energy density as the minimum energy density in nature \cite{Lake:2018zeg}. 
However, we may also generalise canonical QM to include the effects of the smeared background by mapping {\it only} classical coordinates to Hermitian operators, as in canonical quantisation procedures. 

Second, the new interpretation is compatible with the canonical treatment of the weak-gravity limit. 
In the non-relativistic approximation, we must distinguish between contributions to the local gravitational force induced by modifications of the time-time component of the metric, $g_{00}$, and~the space-space part, $g_{ij}$. 
Indeed, in the weak-field limit of classical gravity, the vacuum Einstein field equations reduce to Laplace's equation, $\nabla^2 \sqrt{g_{00}} \simeq 0$ \cite{Hobson:2006se,DiracGR:book}. 
For a point source of mass $m$, $g_{00} = c^2(1- 2Gm/(c^2r))$ so that we recover the familiar Newtonian potential $\Phi = -Gm/r$ from the warping of {\it time} alone. 
(See \cite{Rovelli:time} for an interesting non-technical discussion of this point.) 
Although it is rarely highlighted in introductory texts on general relativity, we note that the {\it spatial} curvature of the background is formally set equal to zero in the Newtonian approximation, i.e., $R_{ijkl} := 0$. 
This allows us to to treat the Newtonian potential as a field on a flat Euclidean geometry and to replace the curved-space radial coordinate $r$ with the Cartesian distance $r = \sqrt{\eta_{ij}x^{i}x^{j}} = \sqrt{x^2+y^2+z^2}$. 
(Strictly, this substitution is not possible in the Schwarzschild geometry, in which global Cartesian coordinates cannot be consistently defined \cite{DiracGR:book}.)

Hence, taking the stricter interpretation of $g$ allows us to ``smear'' the Newtonian potential of canonical non-relativistic gravity, which is given its standard interpretation as a scalar field defined on flat Euclidean space \cite{LandauMechanics,DiracGR:book}. 
In this case, we may use the techniques outlined in \cite{Lake:2018zeg} for the smearing of an arbitrary potential in the generalised Schr{\" o}dinger equation.
Nonetheless, we may consider the restricted smeared-space model, i.e., the model in which $g(\vec{r}{\, '}-\vec{r})$ is interpreted as the probability amplitude for the transition $\vec{r}{\, '} \leftrightarrow \vec{r}$, rather than the more general transition $\vec{r}{\, '} \rightarrow \vec{r}$, as an approximation to a more general model in which genuine spatial curvature fluctuations {\it are} generated. 
This corresponds to the description of stronger gravitational fields in which contributions to the field strength from spatial curvature cannot be neglected. 
However, the construction of such a model lies outside the scope of the present work and is left to future studies. 
Clearly, analogues of the above arguments apply equally well to Euclidean momentum space, and~consistency requires us to reinterpret $\tilde{g}_{\beta}(\vec{p}{\, '}-\vec{p})$ as the probability amplitude for the transition $\vec{p}{\, '} \leftrightarrow \vec{p}$.

\section{Subtleties with Spin in Canonical QM} \label{Appendix.C}

In this appendix, we consider various subtleties that arise in the canonical theory of quantum mechanical spin. 
We focus on issues that are relevant for our proposed generalisation of the theory, presented in Section \ref{Sec.5}. 
For this reason, Sections \ref{Appendix.C.1}--\ref{Appendix.C.3} have analogous structures to Sections \ref{Sec.5.1}--\ref{Sec.5.3} and may be read in parallel, if desired.

\subsection{Historical Development of the Theory} \label{Appendix.C.1}
 
The phenomenon of quantum spin was first discovered empirically via the Zeeman effect \cite{Rae}. 
The splitting of atomic energy levels in the presence of an external magnetic field suggested that electrons possessed a kind of ``internal'' angular momentum, that was able to couple to (i.e., interact with) their quantised orbital angular momentum. 
This was later confirmed, explicitly, by the experiments of Stern and Gerlach \cite{Stern-Gerlach}.  
As a possible mathematical description of this phenomenon, Pauli sought operators that satisfied the angular momentum Lie algebra, Equation (\ref{ll_commutator}), but whose concrete representations contained only constant matrix elements. 
He reasoned that such operators represent the intrinsic (coordinate independent) rather than extrinsic (coordinate dependent) properties of quantum particles \cite{Rae}.

Later, the Pauli matrices for spin-$1/2$ particles, $\left\{\sigma_{i}(1/2)\right\}_{i=1}^{3}$, were identified as representations of the ${\rm SU}(2)$ group generators \cite{GroupTheoryJones}. 
This is the double cover of the rotation symmetry group, ${\rm SO}(3)$, and~shares the same Lie algebra, but its elements admit representations with both integer and half-integer eigenvalues \cite{GroupTheoryJones}. 
These have no classical analogues and describe the internal angular momentum (now called spin) states of two very different types of fundamental particle. 
Particles with integer spin, called bosons, obey Bose--Einstein statistics and are able to condense freely into compact multi-particle states \cite{BECs}. 
In short, identical bosons can ``share space'', since there are no obstructions to the spatial overlap of individual particle wave functions. 
By contrast, particles with half-integer spin, called fermions, obey Fermi-Dirac statistics and cannot condense in this way \cite{DiracQM:book}. 

Formally, the spin-statistics theorem states that it is not possible for two fermions to have the same values of all four quantum numbers: $n$, the principle quantum number, $l$, the azimuthal quantum number, $m_l$, the magnetic quantum number, and~$m_s$, the spin quantum number. 
Since the wave functions of any two fermions with the same four quantum numbers would overlap, the theorem forbids identical fermions from sharing the same region of physical space if the $z$-components of their spins are aligned. 
This is known as the Pauli Exclusion Principle.
Mathematically, it arises from the fact that requiring a many-particle wave function to be single-valued is equivalent to requiring it to be antisymmetric with respect to the exchange of any two particles. 
It follows that bosons occupy symmetric quantum states whereas fermions occupy antisymmetric states \cite{Rae,DiracQM:book}.

Finally, the physical origin of quantum mechanical spin was discovered by Dirac, who showed that it arose as a necessary consequence of combining the principles of quantum theory, expressed via the de Broglie relations $E = \hbar\omega$, $\vec{p} = \hbar\vec{k}$, with the principle of Lorentz invariance, expressed via the relativistic energy-momentum relation, $E^2 = p^2c^2 + m^2c^4$ or $E = \pm \sqrt{p^2c^2 + m^2c^4}$. 
Roughly speaking, although the two forms of the energy-momentum relation are classically equivalent, combining the former with the de Broglie relations leads to the Klein--Gordon equation, while combining these with the latter leads to the Dirac equation \cite{Rae,Peskin:1995ev}. 
The first describes the dynamics of spin-$1/2$ fermions whereas the second describes the dynamics of bosons and is obeyed by all free quantum fields \cite{Peskin:1995ev}. 
The Dirac equation is manifestly invariant under ${\rm SU}(2)$ symmetry and is expressed in terms of the Dirac gamma matrices, $\left\{\gamma^{\mu}\right\}_{\mu=0}^{3}$. 
These, in turn, can be expressed in terms of the spin-1/2 Pauli matrices and the two-dimensional identity matrix, $\mathbb{1}_{2}$ \cite{Rae,Peskin:1995ev}. 

In Section \ref{Appendix.C.2}, we review the structure of the Pauli matrices and their associated algebras in more detail, highlighting the difference between representations with integer and half-integer spin. 
We then review the structure of the canonical gamma matrices and the Dirac equation in Section \ref{Appendix.C.3}. 
Our analysis of relativistic fermions is restricted to the treatment of spin-$1/2$ particles, since no fundamental particles with higher half-integer spin values are known to exist in nature \cite{Peskin:1995ev}. 

\subsection{Algebra and Uncertainty Relations} \label{Appendix.C.2}

For $s = 1/2$, the Pauli matrices are
\begin{eqnarray} \label{Pauli_matrices_1/2}
\sigma_{1} = 
\begin{bmatrix}
    0  &  1  \\
    1  &  0 
\end{bmatrix}
\, , \, \sigma_{2} =
\begin{bmatrix}
    0  &  -i \\
     i  &  0 
\end{bmatrix}
\, , \, \sigma_{3} =
\begin{bmatrix}
    1  &  0  \\
    0  &  -1 
\end{bmatrix}
\, ,
\end{eqnarray}
where $\sigma_1 = \sigma_{x}$, $\sigma_2 = \sigma_{y}$ and $\sigma_3 = \sigma_{z}$, by convention. 
These form the fundamental representation of the ${\rm SU}(2)$ group generators, but the Pauli matrices for all higher-order spins can be obtained, straightforwardly, using Kramer's method \cite{KramersMethod}. 
For arbitrary spin, $s$, the corresponding generators are $(2s+1)$-dimensional square matrices. 

The Pauli matrices for all spin values satisfy the ${\rm su}(2)$ Lie algebra:
\begin{eqnarray} \label{sigma_commutator}
[\sigma_{i}, \sigma_{j}] = 2 i\epsilon_{ij}{}^{k}\sigma_{k} \, .
\end{eqnarray}
For the spin-$1/2$ representation, this follows from the identity:
\begin{eqnarray} \label{fundamental_relation_Pauli_matrices}
\sigma_{i}\sigma_{j} = \delta_{ij} \mathbb{1} + i\epsilon_{ij}{}^{k}\sigma_{k} \, . 
\end{eqnarray}
However, the Pauli matrices for other spin values do not satisfy this relation. 
Equations (\ref{fundamental_relation_Pauli_matrices}) imply both the canonical commutation relations (\ref{sigma_commutator}) and the canonical anti-commutation relations:
\begin{eqnarray} \label{sigma_anti-commutator}
[\sigma_{i}, \sigma_{j}]_{+} = 2\delta_{ij} \mathbb{1} \, , 
\end{eqnarray}
also known as the ${\rm SU}(2)$ Clifford algebra \cite{CliffordAlgebras}. 
We stress that the Pauli matrices for spin-$1/2$ fermions satisfy both the Lie and Clifford algebras, Equations (\ref{sigma_commutator}) and (\ref{sigma_anti-commutator}), whereas those for other spin values satisfy only the Lie algebra (\ref{sigma_commutator}). 

For any spin, $s$, the canonical spin-measurement operators are related to the corresponding Pauli matrices via:
\begin{eqnarray} \label{s}
\hat{s}_{i} = \frac{\hbar}{2} \, \sigma_{i} \, . 
\end{eqnarray}
These have $(2s+1)$ eigenvectors, corresponding to the eigenvalues $-s\hbar, -(s-1)\hbar \dots (s-1)\hbar, s\hbar$, and~obey the commutation relations
\begin{eqnarray} \label{ss_commutator}
[\hat{s}_{i},\hat{s}_{j}] = i\hbar \epsilon_{ij}^{k} \, \hat{s}_{k} \, .
\end{eqnarray}
For spin-$1/2$ particles, the spin operators also obey the relation
\begin{eqnarray} \label{fundamental_relation_spin_ops}
\hat{s}_{i}\hat{s}_{j} = \left(\frac{\hbar}{2}\right)^2 \, \delta_{ij} \hat{\mathbb{1}} + \, i\left(\frac{\hbar}{2}\right)\epsilon_{ij}{}^{k}\hat{s}_{k} \, , 
\end{eqnarray}
and, hence, the anti-commutation relations
\begin{eqnarray} \label{ss_anti-commutator}
[\hat{s}_{i},\hat{s}_{j}]_{+} = \frac{\hbar^2}{2} \, \delta_{ij} \, \hat{\mathbb{1}} \, .
\end{eqnarray}

The total spin operator $\hat{s}^2$ is the ${\rm SU}(2)$ Casimir operator (scaled by $\hbar^2)$ and the associated Casimir invariant is $s(s+1)\hbar^2$, giving \cite{Rae,LieAlgebrasGutowski}:
\begin{eqnarray} \label{s^2}
\hat{s}^2 = \left(\frac{\hbar}{2}\right)^2\sum_{i=1}^3 \sigma_{i}^2 = s(s+1)\hbar^2 \hat{\mathbb{1}} \, .
\end{eqnarray}
It follows that $\hat{s}^2$ commutes with all group generators, 
\begin{eqnarray} \label{}
[\hat{s}^2,\hat{s}_i] = 0 \, , 
\end{eqnarray}
so that the simultaneous eigenvectors of $\hat{s}^2$ and $\hat{s}_z$ are chosen (again, by convention) as the basis vectors for the spin Hilbert space \cite{Rae}.  
For $s=1/2$, these are
\begin{eqnarray} \label{}
\Big|\frac{3\hbar^2}{4}, \frac{\hbar}{2}\Big\rangle = \ket{\uparrow} \, , \quad \Big|\frac{3\hbar^2}{4}, -\frac{\hbar}{2}\Big\rangle = \ket{\downarrow} \, , 
\end{eqnarray}
and Equation (\ref{s^2}) is satisfied by the fact that the matrices $\left\{\sigma_{i}(1/2)\right\}_{i=1}^{3}$ (\ref{Pauli_matrices_1/2}) are involutions, i.e.,
\begin{eqnarray} \label{}
\sigma_{1}^2(1/2) = \sigma_{2}^2(1/2) = \sigma_{3}^2(1/2) = \mathbb{1}_2 \, .
\end{eqnarray}

\subsection{Relativistic Spin and the Gamma Matrices} \label{Appendix.C.3}

In $(3+1)$-dimensional Minkowski space, relativistic spin-$1/2$ fermions are described by the Dirac equation,
\begin{eqnarray} \label{Dirac_eqn}
i\hbar \gamma^{\mu}\partial_{\mu}\psi - mc \psi = 0 \, ,
\end{eqnarray}
where $\left\{\gamma^{\mu}\right\}_{\mu=0}^{3}$ are the Dirac gamma matrices. 
In the Weyl, or chiral, representation these are given by \cite{Peskin:1995ev}:
\begin{eqnarray} \label{gamma_matrices-Dirac}
\gamma^{0} = 
\begin{bmatrix}
    0  &  \mathbb{1}_2 \\
    \mathbb{1}_2  & 0 
\end{bmatrix}
\, , \quad
\gamma^{i} = 
\begin{bmatrix}
    0  &  \sigma_{i}  \\
    -\sigma_{i}  &  0 
\end{bmatrix}
\, ,
\end{eqnarray}

It is straightforward to show that the gamma matrices (\ref{gamma_matrices-Dirac}) satisfy the canonical anti-commutation relations:
\begin{eqnarray} \label{gamma_anti-commutator}
[\gamma^{\mu},\gamma^{\nu}]_{+} = 2\eta^{\mu\nu} \mathbb{1}_4 \, , 
\end{eqnarray}
where $\eta_{\mu\nu}$ is the Minkowski metric, expressed in terms of the ``Cartesian'' space-time coordinates $\left\{t,x,y,z\right\}$. 

\section{Philosophical Issues with the Graviton: Quantum of Space-Time or Quantum of Curvature?} \label{Appendix.D}

In this appendix, we discuss the physics of gravitons in more detail. 
In particular, we focus on how existing results in quantum gravity theory relate to the results of our present work, which suggest that the fundamental quanta of space-time are {\it fermionic} in nature.

We begin by noting that, according to general relativity, classical gravity is the curvature of space-time \cite{Hobson:2006se}. 
Hence, space-time exists even when gravity does not. 
This assumption is the conceptual basis of special relativity and a cornerstone of quantum field theory, which is the quantum theory of matter in flat space-time \cite{SRFrench,Peskin:1995ev}. 
The non-relativistic limits of these theories yield Newtonian mechanics and canonical QM, respectively, both of which are formulated in flat Euclidean space \cite{LandauMechanics,Rae}. 

Thus, in any classical or canonical quantum theory, a background geometry exists, even if it is not curved. 
In relativistic theories, where the gravitational field {\it is} the curvature of the background, assuming zero curvature is equivalent to switching off gravity.  
This is the case for the Standard Model of particle physics, which is formulated in the limit $G \rightarrow 0$.

In the Newtonian approximation, however, this is no longer true. 
As discussed previously in Appendix  \ref{Appendix.B}, taking the static weak-field limit of the vacuum Einstein equations yields the familiar Laplace equation, in which the role of the gravitational potential is played by the time-time component of the metric: $\nabla^2\sqrt{g_{00}} \simeq 0$, $g_{00} = c^2(1 + 2\Phi/c^2)$ \cite{Hobson:2006se,DiracGR:book}. 
Newtonian gravity is therefore due, solely, to the warping of time. 
In this limit, the Riemmanian curvature of the spatial part of the metric has a negligible effect on the total gravitational field strength and, formally, is set equal to zero. 
It is this fact that allows us to describe the Newtonian gravitational potential as a scalar field on a flat Euclidean background.  

This is a subtle but important aspect of the procedure used to recover the Newtonian limit from general relativity. 
It is important, for our purposes, since ``gravitons'' are usually identified with quantised perturbations of the metric \cite{Crowell:2005ax}. 
In the standard treatment, due originally to Pauli and Fierz \cite{Pauli-Fierz}, this is decomposed as:
\begin{eqnarray} \label{classical_pert}
g_{\mu\nu} = \eta_{\mu\nu} + h_{\mu\nu} \, ,
\end{eqnarray}
where $\eta_{\mu\nu}$ is the Minkowski metric and $|h_{\mu\nu}| \ll 1$. 
Crucially, it is assumed that the Minkowski piece remains classical and only $h_{\mu\nu}$ is quantised:
\begin{eqnarray} \label{quantised_pert}
h_{\mu\nu} \mapsto \hat{h}_{\mu\nu} \, . 
\end{eqnarray}
The quantised perturbations $\hat{h}_{\mu\nu}$ obey the Pauli--Fierz equations, which describe the dynamics of spin-2 particles in flat Minkowski space \cite{Pauli-Fierz}:
\begin{eqnarray} \label{Pauli-Fierz}
\Box \hat{h}_{\mu\nu} = 0 \, , \quad \partial_{\nu} \hat{h}_{\mu\nu} = 0 \, , \quad \hat{h}^{\mu}{}_{\mu} = 0 \, .
\end{eqnarray}
Hence, {\it gravitons are bosons}.

Clearly, we may recover the classical Newtonian limit from Equation (\ref{classical_pert}) by setting $h_{ij} = 0$ and $h_{00} \propto \Phi$. 
In principle, we may also recover the ``quantum'' Newtonian limit from Equations (\ref{classical_pert})--(\ref{Pauli-Fierz}) by performing the same classical approximations before mapping $h_{00} \mapsto \hat{h}_{00} \propto \hat{\Phi}$.
However, one may ask the question: what is the quantum description of the background space when gravity is negligible? 

Put another way, if gravitons describe quantised curvature (i.e., the quantised gravitational field), of what is the quantum space-time composed when it is flat? 
In this case there is no spatial curvature and no warping of time. 
It stands to reason that the resulting space-time cannot be composed of gravitons. 

Thankfully, established physics suggests an answer to this question. 
In quantum field theories, forces are mediated by virtual bosons, and~the fundamental particles that ``feel'' these forces are real fermions \cite{Peskin:1995ev}. 
It is therefore possible, and~perhaps even likely, that the fundamental quanta of space-time are fermionic in nature. 
The exchange of virtual spin-2 bosons (gravitons) between space-time quanta could then describe quantised curvature, by analogy with the gauge field description of other fundamental forces. 
With this in mind, we recall that field strengths can also be interpreted as curvatures in the fibre bundle approach to gauge theories \cite{Nakahara:2003nw}. 
For gravity, the extreme nonlinearity of the field equations implies that virtual gravitons are self-interacting \cite{Garay:1994en}, as opposed to say, virtual photons \cite{Peskin:1995ev}, which introduces many complications in the quantum regime. 
However, there is no reason why gravitons should, even in principle, be the fundamental quanta of space-time, as well as of space-time curvature. 

In a more fundamental scenario, both the flat space-time of specially relativistic theories, and~the flat Euclidean space of non-relativistic models, should admit quantum descriptions. 
The smearing procedure, proposed in \cite{Lake:2018zeg} and extended in the present work, is intended as a first step towards their construction. 
As shown in Section \ref{Sec.5.2}, the consistent description of angular momentum and spin in the smeared-space model {\it requires} the quantum state associated with the background space to be fermionic. 
However, based on the arguments considered here, this does not put it in conflict with the known physics of gravitons.


\end{document}